\documentstyle[11pt,epsf,mssymb]{article}

\begin{document}

\begin {titlepage}
\title{Field Theory reformulated without Self-energy Parts. \\ The dressing
Operator}

\author{Michel {\sc de Haan}}
\author {M. de Haan \\
Service de Physique Th\'eorique et Math\'ematique\\Universit\'e
libre de Bruxelles\thanks{Campus Plaine
CP 231, Boulevard du Triomphe, 1050
Bruxelles, Belgique. email: mdehaan@ulb.ac.be}, Brussels, Belgium. }

\maketitle

\begin{abstract}
{The reformulation of field theory for avoiding  self-energy parts in the dynamical evolution has
been applied successfully in the framework of the Lee  model,\cite{dH03a}\ enabling a kinetic
extension of the description.
The basic ingredient is the recognition of these self-energy parts.\cite{dHG00a}\
The original reversible description is embedded in the new one and appears now as a restricted class
of initial conditions.\cite{dHG03}\
This program is realized here in the reduced formalism
for a scalar field, interacting with a two-level atom, beyond the usual rotating
wave approximation. 
The kinetic evolution operator,  previously surmised,\cite{dH91} is here derived from first
principles, justifying the usual practice in optics
where the common use of the so-called pole approximation\cite{CT94} 
should no longer be viewed as an approximation
but as an alternative description in the appropriate formalism.
That model illustrates how some dressing of the atomic levels (and vertices), through an
appropriate operator,  finds its place  naturally into the new formalism since the bare and dressed
ground states do no longer coincide.
Moreover, finite velocity for field propagation is now possible in all cases, without the presence of
precursors for multiple detections.}
\bigskip


\end{abstract}


\end{titlepage}

\def\theequation{
\thesection.\arabic{equation}}

\section {Introduction}
Quantum optics is not an {\it ab initio} theory and requires  a quantum modelization of the
interacting atoms and fields.
The natural point of view starts with an atom described by its energy levels and a dipolar 
interaction  with the field, in a time reversible formalism, using  hermitian
Hamiltonians.\cite{Mo75}\ 
Moreover, a welcome simplifying approximation (the so called rotating
wave approximation) provides often an excellent approximation of the dynamics involved.
Nevertheless, some problems subsists: they involve the description of the instability of the excited
levels and the acausal behaviour in energy transfer between atoms \cite{PT97} or precursors in the
case of a double photodetection.\cite{dH85}\
The first problem is tackled by the use of irreversible elements inside a reversible quantum
mechanical description.
Indeed, for instance, the explicit attribution inside the Hamiltonian 
of a lifetime to an atomic excited level
is the natural way to take into account its unstability.\cite{CT94}\
The introduction of elements of a  phenomenological origin
into a microscopic description
has then led to tremendous success in that field.
The problems linked with the non-hermiticity are avoided by
skilled use of the formalism.
That practice has so far found no fundamental justification.

In a previous paper,\cite{dH91}  
we have analysed in details a renowned paper \cite{Mo75}
and we have shown that in fact, Mollow's approximations were equivalent 
to the use of kinetic equations 
to provide the description of the system.
Implicitly, such kinetic equations are naturally 
at the level of reduced distribution functions for the field.
Indeed, when an arbitrary number of photons can be emitted
from any initial state (except of course the true ground state),
the most adequate description is a reduced formalism,
applied currently for describing atoms in a fluid 
where quantal reduced distribution functions obey the Bogoliubov-Born-Green-Kirkwood-Yvon
hierarchy
\cite{Bal75} in place of the Liouville-von Neumann equation.

The use of those (kinetic-like) equations in ordinary quantum mechanics
can be criticized on two bases.
The first one is that they do not belong to the (reversible) framework 
of ordinary quantum mechanics and their theoretical justification 
is still missing (despite their extraordinary success).
The second one is that they are always presented 
as the result of a largely justified approximation
and not as an intrinsic property of the system.
It seems unsatisfactory that a radical change of the formalism
(a transition  from a time reversal invariant to an irreversible one)
results  simply from approximations.

Physicists associate the concept of an unstable state to an object that, outside external influence,
would decay in a purely exponential way.
Such an unstable state is beyond the reach  of ordinary quantum mechanics and the tentatives to
define it make trouble for normalization properties for instance.\cite{NN58}\
The previous attempts for obtaining a purely exponential decay \cite{dHH73a} were unsatisfactory
since they did not incorporate the possibility of an excitation  mechanism.
The main aim of this paper is to show that an association between intuitive concepts and quantal
description is possible, through the single subdynamics approach, for the interaction  of a field with
a two-level atom, including the consideration of the counter-rotating terms, keeping in mind the
generalization to multilevel systems.
Acausal behaviours can moreover be excluded by an appropriate choice of a dressing operator.
Under compatibility conditions to be satisfied at initial times, it is possible to show that
the time reversal invariant and kinetic  descriptions bring simply different aspects to light while
keeping their equivalence (A spectral representation shares also that property. The derivation of
a spectral-like representation for the Lee model through our single subdynamics approach can be
found in Ref. \cite{dHG04}).

A theory of  subdynamics has been introduced thirty years ago by the Brussels group (see e.g.
\cite{PGH69}, \cite{Bal75}) for a dynamics provided by the Liouville-von Neuman equation.
In that quest for the introduction of irreversibility  inside the formalization of dynamics,
the subdynamics concept  has been shown to be fruitful.
Different realizations are possible according to a choice
of the vacuum, i. e. the choice of the degrees of freedom
that are included in the resulting dynamics, 
while the other elements of the description
become functional of the vacuum ones.
A setback of that approach is thus a limitation on the class of possible initial conditions since they
have to belong to the subdynamics.
Therefore, the initial formulation should be in some sense overcomplete and contain degrees on
freedom on which no control is possible. 
A way out has been the introduction of a {\it transformation theory}, intensively
studied.\cite{PGHR73}\
 In the original ambition, an association of a so-called {\it physical representation} with real
(energy-conserving) processes between renormalized (dressed) quantitities.
It could not be carried out in a general and consistent fashion because of difficulties that have been
reviewed.\cite{dH98}\ 
Moreover, the problematics of ensuring positivity of the density matrix is
lacking in late papers 
\cite{OPP01} where the positive character of the density matrices is no more listed among the
requirements on the $\Lambda$ transformation.

 In order to keep the completeness of the description,
a new vacuum concept has been introduced \cite{dHG00a} for models in field theory, irrespectively
of their classical \cite{dH04b} or quantum character.  
It is based on a dynamical analysis of all
possible contributions  to the formal solution of the Liouville-von Neumann equations.
To avoid the previous trap, the so-called single subdynamics approach \cite{dHG00a} is based on
the existence of self-energy contributions to the dynamics.
We can indeed accept that no control is possible on these processes.
By definition, all self-energy parts have to be excluded from the vacuum.
Their recognition implies that 
the initial dynamics is first extended by discriminating among
the degrees of freedom according to their status 
with respect to preparation and observation.
A subdynamics of the (extended) dynamics is then introduced 
such that it encompasses the original dynamics 
but does no longer contain dressing processes.
In that way, we obtain a reformulation of field theory that excludes self-energy contributions in the
dynamics. 
They are now driven by the other degrees of freedom and provided by time independent functionals
of the other degrees of freedom that are the motor of the evolution.
Therefore, the same mathematical tool (subdynamics) as the Brussels-Austin group is  used, 
but with a different realization, leading to different physical content, although a similar aim is
pursued. 

In a previous paper in collaboration with C. George,\cite{dHG03} we have been dealing with
the Friedrichs model, equivalent to a specific sector of a
two-level atom interacting with a scalar field within the rotating wave approximation (RWA). 
The Friedrichs-Lee model 
has been treated in two different approaches.
In the first one,\cite{dHG03} the existence of sectors has been used to 
perform the explicit construction 
of all the elements of the subdynamics super-operator 
inside the first non trivial sector.
We have shown that a kinetic description exists that
provides an exact and complete alternative 
to the (time reversal invariant) Schr\"odinger description.
It is obtained by a double operation: An enlargement of dynamics, 
that enables the recognition of self-energy parts,
followed by the use of an appropriate subdynamics projector. 
The resolution of the model enables the explicit verification of all the claimed properties.
That proof is welcome since the very existence of the concept of subdynamics 
has been questioned for instance by P. Coveney and O. Penrose.\cite{CP92}\
Their argument is the incompatibility between branch points (generating ``long time tails") and 
the kinetic description, preventing the subdynamics to provide the asymptotic behaviour. 
A general analysis of the situation, leading to their refutation,  can be found in a previous
paper.\cite{dH98}\  
In the second approach,\cite{dH03a} the reduced formalism has been used 
to go beyond the limitation induced by the equivalence with the original description.
The focus has been on the  subdynamics evolution super-operator and we have shown the possibility
of extending quantum theory in a satisfactory way. 
Positivity and normalization are automatically ensured in the  new kinetic description, starting from
the reduced formalism.
Such a construction rests entirely on the existence of poles, independent of branch
points, for diagonal matrix elements of the Green's function associated with the Hamiltonian. 
For this model, a clear-cut separation of  poles and cuts does exist, leading to an intuitive
description and justifying, through a derivation, a phenomenological approach.\cite{dH91}\

To show the robustness of that intuitive description, we treat here a more general system,   
consisting of a two-level atom in interaction
with a scalar field, keeping the counter-rotating terms.
The system under study is interesting from a physical point of view  
since the existence of long time tails (non exponential contributions)
and deviations from exponential behaviours for very short times
has long been recognized in it. 
On the other hand, physicists analysing experiments in optics
are accustomed to use the so-called poles approximations.
How can those empirical rules be justified from first principles?
Are they valid only in some approximate way or do they fully reflect the physical reality.

The associated subdynamics has to be constructed for the new system, outside the RWA. 
Such an extension involves formal modifications in the treatment, but the philosophy is the
same. 
First of all, an analysis of the property of the Liouvillian using sectors \cite{dHG03} is no
longer valid:  A ``reduced formalism" is therefore naturally required.
All sectors are now coupled and have  to be considered together.
As a consequence, the possibility of a complete resolution of the model is lost with respect to the
RWA case.

From the expression of vacuum to vacuum  elements of the resolvent of the generalized
Liouvillian,\cite{dH03b}  we have investigated whether a pole can be associated with all the matrix
elements  that do not involve neither an incoming nor an outgoing field particle,
the generalization of the property for other kinds of matrix elements 
being straightforward.
We have proven \cite{dH03b} that the notion of poles associated with 
the stable and unstable states is still relevant for the model under consideration treated inside the
reduced formalism.

The construction of the evolution and projection super-operator for a subdynamics                                
rests on an analysis of the kind of behaviour for the contribution 
of each term in a perturbation expansion of the resolvent 
(associated with the generator of motion) 
and the selection of the relevant behaviour. 

When compatibility (or equivalence) conditions are satisfied,\cite{dHG03}  the kinetic description
provides somehow a change of representation,  analog to a change of basis in standard quantum
mechanics. Therefore, it will always be possible to transfer the information 
on the system from the usual density matrix operator 
(or its analog for the reduced description)
to the kinetic description and vice versa.
Moreover, the intuitive way of looking at the system is recovered in the kinetic description in terms
of incident field, outgoing field, dressed atomic levels 
(including their attribute of unstability represented by their lifetime,
see Ref. \cite{Mo75} for instance).  
Therefore, the common practice in optics will be justified 
from first principle and will no longer be the result of approximations.
Let us underline the analogy of the present approach  with the theory of renormalization.
Indeed, in both theories, the aim is to take properly into account 
the self-energy contributions.
The renormalization theory tackles the problem of removing ultraviolet divergences 
at the level of the wave function and to derive finite corrections.\footnote{It has been recently
noticed \cite{dH04b} that the application of the single subdynamics approach to a well adapted
formalism of classical electrodynamics \cite{BP74} leads directly to a divergence-free description,
without any need to a substraction procedure.}   
The description of an unstable state requires the introduction of an imaginary part in the
renormalized energy:  
The original framework cannot be respected. In the present theory, in terms of a reduced density
operator, no such problem arises.

In this effective realization  of our approach on a less simple non trivial example,
we will not bother about the formal properties of the subdynamics super-operator
that have been well established \cite{Bal75} and do not depend on the particular realization of the
choice of the vacuum. 
We will not worry either of the compatibility conditions: we know that they do exist  \cite{dHG03}
and reflect the possible equivalence between the original and the kinetic descriptions.
We focus on the derivation  and the properties of the kinetic evolution super-operator,
 as  defined by the single subdynamics approach.\cite{dHG00a}\
The form of the obtained evolution generator leads us naturally to the introdution of
dressing  technique introduced long ago into the Brussels approach.\cite{PGHR73,dHH73a}\
A dressing operator enables to shape the form of the evolution generator corresponding to an
intuitive description, out of the reach of the original time reversal invariant description, that
accepts only unitary transformations.

In  Ref. \cite{dH91}, a reduced formalism has been proposed 
to treat the interaction of a two-level atom with the electromagnetic field.
Various kinetic equations have been accordingly surmised and 
justified on physical grounds.
However, their derivation from first principles 
was outside the scope of that paper
and we intend here to fulfill that missing part:
In this way, we pursue the progressive introduction of the characteristics of the method
to the effective description of the interaction 
between the electromagnetic fields and atoms.

In $\S$2, the reduced formalism for the system is briefly recalled
and the extension of the dynamics 
(distinction between the various photons) is treated.
The formal properties of the subdynamics super-operator are briefly cited.

$\S$3 is devoted to the elements of the  subdynamics super-operator
that enable the obtention of the kinetic operator for the elements
describing the population of the atomic levels.
They do not involve physical photons 
The elements for the atomic dipolar moments are considered in $\S$4.

The main difference with respect to RWA
is already apparent in the elements of the evolution generator $\tilde{\bar\theta}$
that do not involve photons 
(the purely atomic part of the evolution operator).
Inside  the RWA, the bare and dressed ground state coincide and the dressed excited state is
then derived directly by the construction of the kinetic operator.
Outside the RWA, the structure of the kinetic operator is no longer the same: 
both states that appear in it are susceptible of evolution. 
Indeed, already in usual quantum mechanics, we know that
a change of basis is required with respect to the bare states outside RWA.
Therefore, the stable ground state does not coincide with the state described by the kinetic
operator.
On physical grounds, it is required  the true ground state being time independent and the excited
state decaying.
Since we are dealing in a reduced formalism,  we have to translate such properties  into similar ones
for the matrix elements of the reduced density operator, in order to define a physical
representation.
 
Through a dressing operator, usual in the subdynamics approach,
\cite{PGHR73,dHH73a}\ 
a procedure enables to fix the problem in $\S$5 in a completely satisfactory way.
After dressing, the structures of the kinetic operators inside and outside the RWA are common.
Let us note that the atomic model under consideration  is the first example treated
for which a dressing operator is required.
Our dressing, in the single subdynamics approach formulated in reduced formalism, is not
equivalent  of defining new states in an Hilbert space formulation, such as in \cite{SCG78}.

For completeness, the (somehow trivial) effect of passive photons
is explicitly treated in $\S$6.
The one physical photon absorption process is also considered
in  the same $\S$6 and leads, as in the case of the Lee model,\cite{dH03a}
to the introduction of dressed interactions between the atom and the field.
The one physical photon emission process does not present unexpected new features.

The vertex dressing is considered in  $\S$7.
We discuss the change in the interaction that can be induced by the dressing operator:
Strict causality in the exchange of photons between atoms can be  ensured in all cases, without the
usual presence of precursors due to a finite lower bound in the energy
spectrum.\cite{GH74}-\cite{GH98}\ 
The use of a kinetic description, as opposed to a reversible one,
is a main ingredient to allow that property, beyond the reach of an hermitian generator of motion. 

Concluding remarks are presented in the last part of this paper.

\section {The model - Reduction - Indiscernability -\newline Extended Dynamics - Subdynamics}
\setcounter{equation}{0}

The model considered is a two-level atomic system interacting with 
a field, without considering the rotating wave approximation
nor explicitly the momentum change due the recoil of the atom.
The Hamiltonian of the system can be written as
\begin{eqnarray}
H&=&\omega_1a^+_1a_1+\omega_0a^+_0a_0+\sum_k \omega_ka^+_ka_k
\nonumber\\
&+&\sum_k \left(V_{1|0k}a^+_1 a_0 (a_k+a^+_k) 
+V_{0k|1}a^+_0 a_1 (a_k+a^+_k)\right).
\label{2.1}
\end{eqnarray}
In another paper,\cite{dH03b} we have established the analytical properties 
of some Green's functions associated with the Hamiltonian (\ref{2.1}).
Since the Hamiltonian enables an arbitrary number of field particles 
being present in the future of any state,
due to the counter-rotating terms,
we are in a situation similar to that met in statistical mechanics and
a reduced formalism is required for the description 
of the degrees of freedom associated with the field. 
In the computation of the mean value of observables, such a formalism replaces 
the trace operation by a vacuum expectation value for the field.
It has been developed in extenso in Ref. \cite{dH91} and recalled in Ref. \cite{dH03a}
and we will be satisfied with a reminder of the main features
without entering into details.

The many body system of interest is described by a density operator $\rho$
that obeys the Liouville-von Neuman equation
\begin{equation}
i\partial _t \rho=L\rho,
\label{2.2}
\end{equation}  
where the Liouvillian $L$ is the commutator with the Hamiltonian $H$
given by (\ref{2.1}).
In terms of factorizable superoperators ($A\times B$),  defined as
$(A\times B )\rho \equiv A\rho B$,
it is given by
\begin{equation}
L=H\times I-I\times H,
\label{2.3}
\end{equation}  
where $I$ is the identity operator. 
A (factorizable) superoperator $(A\times B)$ will be called a 
connecting superoperator if both $A$ and $B$ are different 
from the identity operator.
A superoperator  such as in $L$ 
that acts with an operator on one side of the density operator
and with the identity operator on the other side will be called  a non-connecting
superoperator.
It does not prevent from writing the formal solution of equation (\ref{2.1}) for $\rho$
under a factorized form $(\exp{-iHt})\,\rho\, \exp{iHt}$.

The associated reduced density operator  $\bar\rho$ obeys an equation
\footnote{We keep the notations of Ref. \cite{dH03a}
where the operators inside the reduction procedure bears a
bar accent \cite{dH91} 
while the symbol tilde is introduced to refer to the extended dynamics.
With respect to Ref. \cite{dH91}, the reduced density operator is noted $\bar\rho$ in place of
$\sigma$.}
\begin{equation}
i\partial_t \bar\rho=\bar L\bar\rho.
\label{2.4}
\end{equation}  
The unperturbed part  $\bar L_0$  of $\bar L$ is the same 
as the unperturbed part  $L_0$ of $L$.
The potential dependent part $\bar L_V$ of $\bar L$ contains, 
in addition to the potential dependent part $L_V$, 
new  connecting contributions $\bar L'_V$
given by
\begin{eqnarray}
\bar  L'_V&=&
\sum_k \left(V_{1|0k}a^+_1 a_0 
+V_{0k|1}a^+_0 a_1 \right)\times a^+_k
\nonumber\\
&-&\sum_k  a_k\times\left(V_{1|0k}a^+_1 a_0 
+V_{0k|1}a^+_0 a_1 \right).
\label{2.5}
\end{eqnarray}  
If we note by a Roman letter the states of the Hilbert space
(that letter defines the state of the atom and the wave numbers
associated with the photons present)
and if we use the notation
$<a|\rho|b>=\rho_{ab}$, the evolution equations take the form:
\begin{equation}
i\partial _t \rho_{ab}(t)=\sum_{c,d}L_{ab.cd}\rho_{cd}(t), \qquad
i\partial_ t {\bar\rho}_{ab}(t)=\sum_{c,d} \bar L_{ab.cd}\bar\rho_{cd}(t)
\label{2.7}
\end{equation}  
with an obvious definition for the matrix elements of the evolution super-operators
$L$ and $\bar L$. 

A formal solution of these equations is provided with the help of inverse
Laplace transform as
\begin{eqnarray}
\rho_{ab}(t)&=&\sum_{c,d}\frac {-1}{2\pi i}\int_\gamma dz \, e^{-izt}
\left(\frac 1{z-L}\right)_{ab.cd}\rho_{cd}(0),
\label{2.8}
\\
{\bar\rho}_{ab}(t)&=&\sum_{c,d}\frac {-1}{2\pi i}\int_\gamma dz \, e^{-izt}
\left(\frac 1{z-\bar L}\right)_{ab.cd}\bar\rho_{cd}(0),
\label{2.9}
\end{eqnarray}  
where the path $\gamma$ lies above the real axis.
These forms enable immediately perturbation expansions in terms of the potential
$V$.
In the first part of this study, 
we limit ourselves to the case of the couples $(ab)$
and $(cd)$ referring to diagonal discrete states without field particles.
The resolvent of $L$ that plays a role in (\ref{2.8}) can be written as a convolution product
of appropriate resolvents of $H$.  

Analytical properties of the resolvent of $L$ and $\bar L$ in (\ref{2.8}-\ref{2.9})
have been studied extensively in Ref. \cite{dH03b} (in particular the elements
$\left((z- L)^{-1}\right)_{11.11}$, $\left((z- L)^{-1}\right)_{00.00}$), $\left((z-\bar
L)^{-1}\right)_{11.11}$ and
$\left((z-\bar L)^{-1}\right)_{00.00}$).
The main technical point involved was the ability of distinguishing the effects
of the branch points and of the poles required for the construction of the single subdynamics.
In the reduced formalism, the analytic structure of the resolvent of the evolution
super-operator can no longer be examined \cite{dH03b} {\it a priori} via a convolution involving the
Green's functions associated with the Hamiltonian operator and this has far reaching consequences on
the level of the analytic properties.

We recall some of the properties that have been established.

The Green's functions associated with 
the Hamiltonian  (\ref{2.1}) are noted as in the RWA  case \cite{dHG03}
\begin{equation}
\frac{1}{\eta_1(z)}= \left(\frac1{z-H}\right)_{11}, 
\qquad
\frac{1}{\eta_0(z)}= \left(\frac1{z-H}\right)_{00}.
\label{2.3a}
\end{equation}
Inside the RWA, the ${\eta_0(z)}$ function reduces to $(z-\omega_0)$.
``Bar" functions $\bar\eta$ represent Green's functions  associated with the operator $(-H)$ 
\begin{equation}
\frac{1}{\bar\eta_1(z)}= \left(\frac1{z+H}\right)_{11}, 
\qquad
\frac{1}{\bar\eta_0(z)}= \left(\frac1{z+H}\right)_{00},
\label{2.4a}
\end{equation}
so that the relation (\ref{2.8}), for $a=b=c=d=1$ or $a=b=c=d=0$ involves the resolvent
$R_{11.11}(z)$ and $R_{00.00}(z)$ given by
\begin{eqnarray}
R_{11.11}(z)&=&\frac{-1}{2\pi i}\int_{\gamma} du\, 
\frac{1}{\eta_1(u)}\frac{1}{\bar\eta_1(z-u)},
\nonumber\\
R_{00.00}(z)&=&\frac{-1}{2\pi i}\int_{\gamma} du\,
\frac{1}{\eta_0(u)} \frac{1}{\bar\eta_0(z-u)}.
\label{2.5a}
\end{eqnarray}
The $\eta(z)$, $\bar\eta(z)$ functions are required in (\ref{2.5a}) for a positive imaginary part
of their argument ($0<\Im\gamma<\Im z$).
Therefore, those functions will be analytically contined from above 
into the lower half plane $\Im z<0$.
When branch points are met, the cuts will be placed parallelly to
the imaginary axis.
All functions of complex arguments that we shall introduce  share 
that property and we shall dispense them from a upper index ``+" that would
indicate the way the analytical continuations are performed.
As a consequence of that convention, for instance, $\bar\eta_1(z)$
for $\Im z<0$ cannot be computed directly from (\ref{2.4a})
but requires the computation of $\left((z+H)^{-1}\right)_{11} $ 
for $\Im z>0$ and then an analytical continuation.
It is usual to note that fact by writing $\bar\eta^+_1(z)$
but to avoid too cumbersome notations, 
we use an implicit convention to keep the notation $\bar\eta_1(z)$.
The same convention holds for other functions to be introduced soon.

The $\eta^{-1}$ functions are analytic in the upper halfplane $\Im z>0$ and
we first analyse their singularities in the lower halfplane $\Im z<0$.
The properties of the $\bar\eta$ functions are a translation of those of 
the $\eta$ functions.
From the convolution product (\ref{2.5a}), the singularities 
of the resolvent into the lower halfplane $\Im z<0$ are then inferred.

From a perturbation expansion with respect to the interaction $V$, 
the $\eta_1^{-1}$ function (\ref{2.3a}) can be written in general as
$\eta_1^{-1}=\sum_{n=0}^{\infty}((z-H_0)^{-1})_{11}$
$\left(f_1(z)((z-H_0)^{-1}\right)_{11})^n$
where  $f_1(z)$ represents  the sum of all the contributions leading from state``1" to itself 
with a condition of  irreductibility
with respect  to ``1" (all intermediate states implied in the sum
have to be $\not=$ of ``1") : 
\begin{equation} 
f_1(z)=\sum_{m=0}^{\infty}((V((z-H_0)^{-1}))_{irr1}^mV)_{11}.
\label{2.5ab}
\end{equation}
The index $irr1$ recalls the restriction on the intermediate states.   
The diagonal matrix elements of $(z-H_0)^{-1}$ are called propagators.

A visual representation of the contributions can be obtained using a diagrammatic representation,
such as the one developped in appendix B of Ref. \cite{dH85} for the same system. 
It corresponds in the present case to the Feynman diagrams.
We have then trivially
\begin{equation} 
\frac{1}{\eta_1(z)}=\frac{1}{z-\omega_1-f_1(z)},
\qquad
\frac{1}{\eta_0(z)}=\frac{1}{z-\omega_0-f_0(z)}.
\label{2.7a}
\end{equation}

The summations present in the expression of the functions $f$ have to be understood in the infinite
volume limit where, for instance, $\sum_{\bf k} |V_{1|0k}|^2 \dots \to \int_0^{\infty}d\omega\,
v^2(\omega)\dots$ 
We assume that the function $v^2(\omega)$ vanishes as $\omega$ for small values of
its argument (This may be understood as it is composed of a factor $k^2$ 
arising from the jacobian to obtain spherical coordinates and the square of 
a usual factor $\frac1{\sqrt \omega_k}$ arising from the matrix element  of the 
potential.).

The thesis established  in Ref. \cite{dH03b} is the following: 
Under this behaviour of the potential matrix elements $V_{1|0k}$, $V_{0|1k}$, $V_{0|1k}$, $V_{1|0k}$
for small value of the argument $k$ , 
we have the consistency of the following statements 
for the singularities of the functions $\eta$ and $f$.

\noindent
1)The function $(\eta_1(z))^{-1}$ presents a pole at some point 
$z=\omega_1+\zeta$ and a well defined residue ${\cal A}_1$ at that point.

\noindent
2)The function $(\eta_0(z))^{-1}$ presents a pole at some point 
$z=\omega_0+\delta$ and a well defined residue ${\cal A}_0$ at that point. 

\noindent
3)The functions $f_1(z)$ and $f_0(z)$ present logarithmic singularities
at the points $z=\omega_1+\zeta$ and $z=\omega_0+\delta$.

\noindent
4)The function $f_1(z)$ behaves as $(z-\omega_1-\zeta)^3\ln(z-\omega_1-\zeta)$ 
in the vicinity of its singular at  point $z=\omega_1+\zeta$ 
and as $(z-\omega_0-\delta)\ln(z-\omega_0-\delta)$ 
in the vicinity of its singular point $z=\omega_0+\delta$. 

\noindent
5)The function $f_0(z)$ behaves as $(z-\omega_0-\delta)^3\ln(z-\omega_0-\delta)$ 
in the vicinity of its singular point $z=\omega_0+\delta$ and as
$(z-\omega_1-\zeta)\ln(z-\omega_1-\zeta)$ 
in the vicinity of its singular point $z=\omega_1+\zeta$.

\noindent
$\zeta$ has an negative imaginary part while $\delta$ is real.
Inside the RWA \cite{dHG03}, the pole of $(\eta_1(z))^{-1}$ has been
noted $\theta_1$.
Here, we bring to the fore the displacement with respect to the unperturbed value $\omega_1$.
For the ``bar" functions, we have the similar properties: the poles of 
$(\bar\eta_1(z)-1)$ and $(\bar\eta_0(z))^{-1}$ are resp. at $z=-\omega_1+\bar\zeta$ and
$z=-\omega_0+\bar\delta$,  with $\bar\delta=-\delta$, $i\bar\zeta=(i\zeta)^*$. 

The explicit demonstration has required a self-consistent analysis of the various contributions.
For the system we consider here (outside the RWA), 
the previous  proof of the existence of the poles inside the RWA \cite{dHG03}  has been adapted:
the structure of the Hamiltonian Green's functions is no longer as simple
and the diagonal elements of the Green's functions present simultaneously a pole and a 
branch point for a same value in the complex plane $z$,
while poles and branchpoints do not coincide inside the RWA.

An analysis of that  new intertwining has enabled to desantangle it in order 
to be able to make a statement about the existence of poles 
independently of a choice of a particular Riemann sheet.
The problematics to which an answer has been given  is thus the existence of a ``pole approximation" 
\cite{CT94}  beyond the RWA.

From the convolution form, (\ref{2.5a}) it is then established that 
$R_{11.11}(z)$ and $R_{00.00}(z)$ have respectively a pole at $z=\zeta+\bar\zeta$ and $z=0$.
In a similar way, it can be easily established that $R_{10.10}(z)$ and $R_{01.01}(z)$ have simple poles
respectively for $z=\omega_1+\zeta-\omega_0+\bar\delta$ and
$z=\omega_0+\delta-\omega_1+\bar\zeta$.

That analysis could not be reproduced {\it mutatis mutandis}.
Indeed, the new generator of motion in the reduced formalism is no longer
likewise simply connected with the Hamiltonian  since the notion of 
reduction leads outside the hamiltonian formalism.
Therefore,  the existence of ``connecting vertices" \cite{dH03a} 
prevents a direct analysis in terms of a convolution involving 
matrix elements of the Green's function associated with the Hamiltonian. 
Nevertheless, a further analysis has shown that both matrix elements  $\bar R_{11.11}(z)$ and $\bar
R_{00.00}(z)$ of the Green's function associated with the reduced Liouvillian $\bar L$ have  poles
$z=\zeta+\bar\zeta$ and
$z=0$, with well defined residues.\cite{dH03b}\
That result is not unxpected on physical ground: the natural time behaviour of the atom should not be
modified by the reduction  procedure. 
The existence of those poles is a necessary requisite 
for the construction of the subdynamics.

The indiscernability of the field 
and its consequences on the computations has already be treated in Ref. \cite{dH03a} and it has
been noted that they are minimal.
We will not repeat the same considerations.
Indeed, in Ref. \cite{dH91}, care has been taken on that question.

The evolution for field and atom reduced distribution functions
is governed by an evolution operator ${\bar L}$. 
Following the approach initiated with the potential scattering \cite{dHG00b,dHG02}
and the atom in interaction with the field 
inside RWA,\cite{dH03a} 
an extended dynamics is now introduced, 
that rests on a distinction between all degrees of freedom,
once the self-energy parts have been recognized.
We use the same notations as in Ref. \cite{dH03a} , a Roman letter $l$ to represent 
a photon involved in the self-energy contribution, a Greek letter $\mu$
for a photon leaving the atom (emitted photon) and a Greek letter $\lambda$
for an incident photon.
The number of reduced distribution functions to be considered is multiplied
accordingly and the evolution is then governed 
by an evolution operator $\tilde{\bar L}$:
\begin{eqnarray}
\tilde{\bar L}&=&
\omega_1a^+_1a_1 \times I   -I\times \omega_1a^+_1a_1 
+\omega_0a^+_0a_0 \times I -I\times \omega_0a^+_0a_0 
\nonumber\\ &
+&\sum_l \omega_la^+_la_l \times I-I\times \sum_l \omega_la^+_la_l
\nonumber\\ &
+&\sum_{\lambda} \omega_{\lambda}a^+_{\lambda}a_{\lambda} \times I
-I\times \sum_{\lambda} \omega_{\lambda}a^+_{\lambda}a_{\lambda} 
+\sum_{\mu} \omega_{\mu}a^+_{\mu}a_{\mu} \times I
-I\times \sum_{\mu} \omega_{\mu}a^+_{\mu}a_{\mu} 
\nonumber\\&
+&\sum_l \left(V_{1|0l}a^+_1 a_0 (a_l+a^+_l) 
+V_{0l|1}a^+_0 a_1 (a_l+a^+_l)\right)\times I
\nonumber\\&
-&I\times \sum_l \left(V_{1|0l}a^+_1 a_0 (a_l+a^+_l) 
+V_{0l|1}a^+_0 a_1 (a_l+a^+_l)\right)
\nonumber\\&
+&\sum_{\lambda} \left(V_{1|0{\lambda}}a^+_1 a_0 a_{\lambda} 
+V_{0{\lambda}|1}a^+_0 a_1 a_{\lambda}\right)\times I
\nonumber\\&
-&I\times \sum_{\lambda} \left(V_{1|0{\lambda}}a^+_1 a_0 a^+_{\lambda} 
+V_{0{\lambda}|1}a^+_0 a_1 a^+_{\lambda}\right) 
\nonumber\\&
+&\sum_{\mu} \left(V_{1|0{\mu}}a^+_1 a_0 a^+_{\mu} 
+V_{0{\mu}|1}a^+_0 a_1 a^+_{\mu}\right)\times I
\nonumber\\&
-&I\times \sum_{\mu} \left(V_{1|0{\mu}}a^+_1 a_0 a_{\mu} 
+V_{0{\mu}|1}a^+_0 a_1 a_{\mu}\right) 
\nonumber\\&
+&\sum_l \left(V_{1|0l}a^+_1 a_0 
+V_{0l|1}a^+_0 a_1 \right)\times a^+_l
-\sum_l  a_l\times\left(V_{1|0l}a^+_1 a_0 
+V_{0l|1}a^+_0 a_1 \right) 
\nonumber\\&
+&\sum_{\lambda} \left(V_{1|0{\lambda}}a^+_1 a_0 
+V_{0{\lambda}|1}a^+_0 a_1 \right)\times a^+_{\lambda}
-\sum_{\lambda}  a_{\lambda}\times\left(V_{1|0{\lambda}}a^+_1 a_0 
+V_{0{\lambda}|1}a^+_0 a_1 \right)
\nonumber\\&&
\label{2.10}
\end{eqnarray}
The system is now described by a set of new reduced distribution functions $\tilde {\bar\rho}$
related to the extended dynamics.
The constitutive relations connects the original set ${\bar\rho}$ 
to the new one $\tilde {\bar\rho}$.\cite{dHG00a}\
Indeed, the new description contains obviously more degrees of freedom than the original one.
Since the observables are defined originally in terms of ${\bar\rho}$, we have to specify the relation
between the two descriptions.
The constitutive relations do not involve the ``$l$" photons (involved in a self-energy process) and
enable the interference between emitted and incident photons.

The construction of the subdynamics requires 
a classification of the states into two classes, 
the vacuum states and the correlated states.
They are obtained traditionnally \cite{PGHR73}
by the introduction of an  superoperator $P$ 
that projects on the so called vacuum states.
In  our case, correlated states contain at least one intermediate field line 
(of the $l$ type).
Vacuum states contain thus only incoming and outgoing field lines.

The construction rule for the subdynamics operator can be formulated on
the formal solution of the evolution equation in the extended dynamics:
\begin{equation}
\tilde {\bar\rho}_{ab}(t)=\sum_{c,d}\frac {-1}{2\pi i}\int_c dz \, e^{-izt}
\left(\frac 1{z-\tilde{\bar L}}\right)_{ab.cd} \tilde {\bar\rho}_{cd}(0).
\label{2.11}
\end{equation}  
The path $c$ is chosen above the real axis for $t>0$.
Analytic continuation of the integrand from $\Im z>0$  to $\Im z<0$ is
therefore required upon integration over $z$. 
The symmetry with respect to time inversion is thus broken by the procedure.
All integrations on intermediate field lines are (at least)
formally performed to permit an analytical continuation from above of the functions of $z$ so
defined, placing all cuts parallely to the imaginary axis.
The rule is to pick up the contribution of the poles associated with vacuum states, 
The just mentioned analytical continuation from above 
enables to avoid accidental coincidence of the poles associated with the vacuum and correlation
states. 
The physical poles are located at a value defined by the poles associated with the atomic
states, i.e.  $z=\zeta+\bar\zeta$,  $z=0$,
$z=\omega_1+\zeta-\omega_0+\bar\delta$, 
$z=\omega_0+\delta-\omega_1+\bar\zeta$, in addition to frequencies associated with physical
field lines (incoming or outgoing).
For instance, the vacuum-vacuum element
 $\left((z-{\tilde{\bar L})^{-1}}\right)_{1\lambda 0\mu.1\lambda1}$ may have the following physical
poles:
 $z=\zeta+\bar\zeta-\omega_\lambda$,  $z=-\omega_\lambda$,
$z=\omega_1+\zeta-\omega_0+\bar\delta-\omega_\lambda+\omega_\mu$, 
$z=\omega_0+\delta-\omega_1+\bar\zeta-\omega_\lambda+\omega_\mu$.

In the perturbation approach, we have the following expansion for the resolvent $R_{ab.cd}(z)$ of $L$
(involving the resolvent $R^0(z)$ of the unperturbed hamiltonian $H_0$
and the interaction part $L_V$ of the liouvillian):
\begin{equation}
R_{ab.cd}(z)=
\sum_{n=0}^{\infty} \left(R^0(z) \left[L_V R^0(z)\right]^n\right)_{ab.cd}.
\label{2.12}
\end{equation}  
For the resolvent $\tilde{\bar R}_{ab.cd}(z)$ of $\tilde{\bar L}$
we have similarly the expansion:
\begin{equation}
\tilde{\bar R}_{ab.cd}(z)=
\sum_{n=0}^{\infty} \left({\tilde R}^0(z) 
\left[\tilde{\bar L}_V {\tilde R}^0(z)\right]^n\right)_{ab.cd}.
\label{2.13}
\end{equation}
The singularities of the resolvent $R(z)$ of $L$  are defined on resummed expressions. 
Therefore, useful expressions are obtained by considering
irreductible operators with respect to the vacuum. 
For the study of $R_{ab.cd}(z)$, 
where no extension of dynamics has been defined, 
we may define here the set of vacuum states by the states without any field particles.\footnote{In
the more usual approach by the Brussels group, the vacuum has been defined by the set of diagonal
elements or by an adequate extension in the approach 
by the patterns of correlation.\cite{Bal75}}
Such a vacuum is useful since it will provide a point of comparison
for the elements of $\tilde{\bar R}_{ab.cd} $ that do involve 
neither incident nor emitted photons.
The collision operator \cite{PGHR73} $\psi(z)$,
is defined as the sum of irreductible fragments 
leading from a vacuum state to another: 
all intermediate states have to imply at least one field particle: 
\begin{equation}
\psi_{ab.cd}(z)=
\sum_{n=0}^{\infty} \left( \left[L_V R^0(z)\right]^n L_V\right)_{ab.cd
\left(irr\right)}
\label{2.14}
\end{equation}  
where the index $irr$ means that al intermediate states 
involve at least one field line.

We express also the perturbation expansion of the resolvent 
$\tilde{\bar R}(z)$ in terms of irreductible operators: 
For the complete Liouvillian  $\tilde{\bar L}$,
we introduce by analogy the irreductible operators  $W_{ab.cd}$ 
(that could also be noted by $\tilde{\bar \psi}_{ab.cd}(z)$)
\begin{equation}
W_{ab.cd}(z)=
\sum_{n=0}^{\infty} 
\left( \left[\tilde{\bar L}_V 
\tilde{\bar R}^0(z)\right]^n 
\tilde{\bar L}_V\right)_{ab.cd 
\left(irr\right)}
\label{2.15}
\end{equation}  

When $a$, $b$, $c$, $d$ represents states without photons,
the operators $\tilde{\bar R}_{ab.cd}(z)$ and ${\bar R}_{ab.cd}(z)$ 
coincide.
In order to be able to make the connection between the singularities of 
$R(z)$ and $\bar R(z)$, let us define irreductible operators that involve
at least one of the connecting vertices $\bar  L'_V$ (\ref{2.5}) as:
\begin{equation}
T_{ab.cd}(z)=
\left( \left[{\bar L}_V 
{\bar R}^0(z)\right]^n 
{\bar L}_V\right)_{ab.cd 
\left(irr,con\right)}
\label{2.16}
\end{equation}  
where the subscript $con$ implies that at least one of the vertices
${\bar L}$ is  a connecting contribution $\bar L'_V$ (\ref{2.5}).
The link with the above introduced operators $\psi$ and $W$ 
is then obvious:
\begin{equation}
W_{ab.cd}(z)=\psi(z)_{ab.cd}+T(z)_{ab.cd}
\label{2.17}
\end{equation}  
In view of the characteristics of the vacuum states 
(an incoming field line cannot be reintroduced 
and an outgoing field line never disappears),
the contributions can be easily classified 
according to the number of vacuum field lines involved and computed in a recurrent way.
Therefore, we  first consider the contribution of terms 
that do not involve any field line.

\section{Elements of $\tilde{\bar \Sigma}$ without field }                                                                

\setcounter{equation}{0}

Our aim in this section is the computation of the elements of 
subdynamics super-operator $\tilde{\bar \Sigma}_{ab.cd}(t)$ 
when $a$, $b$, $c$, $d$ represent states without photons.
As in the case of the RWA,\cite{dH03a}  
they determine  the evolution super-operator 
$\tilde{\bar \Theta}_{at}$ that does not involve 
absorption nor emission processes.
These elements of  $\tilde{\bar \Sigma}_{ab.cd}(t)$ are computed from
the corresponding elements of the resolvent.
As no physical field lines are involved, 
the extension of dynamics plays no role:
only virtual photons, involved in self-energy contributions, play a role:
We may use the Liouvillian $\bar L$ in place of $\tilde{\bar L}$.   
The elements of the operators $\tilde{\bar R}_{ab.cd}(z)$ and ${\bar R}_{ab.cd}(z)$ coincide and
the analytical properties of ${\bar R}_{ab.cd}(z)$
have been established in Ref. \cite{dH03b} 
for some diagonal-diagonal elements
and the extension of the proof to the other matrix elements 
is straightforward.

The elements of  $\tilde{\bar \Sigma}$ are expressed in terms of
the irreductible operators $W$, $\psi$, $T$.
For the elements of these operators without photons,  some useful relations may be considered.
The last vertex (at the extreme left)
in the perturbative expansion (\ref{2.15}) for $W_{ab.cd}(z)$
corresponds to the absorption of the last virtual photon.
For that vertex, the replacement of the element of $\bar L'_V$ 
by the element of $L_V$ that absorbs also the last photon, or vice versa,
transforms the contribution to $W_{ab.cd}(z)$ 
into a contribution to $W$ with other values of the first two atomic indices, 
leading to the relations:
\begin{eqnarray}
W_{11.cd}+W_{00.cd}&=&0,
\label{3.1}
\\
W_{10.cd}+W_{01.cd}&=&0.
\label{3.2}
\end{eqnarray}  
Since  $\psi_{11.00}(z)$, $\psi_{00.11}(z)$, $\psi_{10.01}(z)$ and $\psi_{01.10}(z)$ vanish,
the links (\ref{2.17}), with the previously introduced operators $\psi$ and $T$,
provides the following relations 
\begin{eqnarray}
T_{11.00}(z)&=& -W_{00.00}(z)=-\psi_{00.00}(z)-T_{00.00}(z),\nonumber\\
T_{00.11}(z)&=&-W_{11.11}(z)=-\psi_{11.11}(z)-T_{11.11}(z),
\label{3.3}
\\
T_{10.01}&=&-W_{01.01}(z)=-\psi_{01.01}-T_{01.01},\nonumber\\
T_{01.10}&=&-W_{10.10}(z)=-\psi_{10.10}-T_{10.10}.
\label{3.4}
\end{eqnarray}
This section is devoted to the elements
$\tilde{\bar \Sigma}_{11.11}$, $\tilde{\bar \Sigma}_{00.00}$, 
$\tilde{\bar \Sigma}_{11.00}$, $\tilde{\bar \Sigma}_{00.11}$
of the subdynamics super-operator $\tilde{\bar \Sigma}(t)$
in order to obtain the elements 
$\tilde{\bar \Theta}_{11.11}$, $\tilde{\bar \Theta}_{00.00}$, 
$\tilde{\bar \Theta}_{11.00}$, $\tilde{\bar \Theta}_{00.11}$
of the evolution operator $\tilde{\bar \Theta}$.
The other elements without physical field lines, such as 
$\tilde{\bar \Sigma}_{10.10}$, are not dynamically connected to the previous
and are treated in the next $\S$4.

The relevant elements of $\tilde{\bar \Sigma}$ 
are the same as those of
$\bar { \Sigma}_{11.11}$, $\bar { \Sigma}_{00.00}$, 
$\bar { \Sigma}_{11.00}$, $\bar { \Sigma}_{00.11}$
computed by choosing the diagonal atomic states as the only elements of the vacuum. 
 In terms of the irreductible operators,
the following relations hold:
\begin{eqnarray}
\bar R_{11.11}(z)&=&R_{11.11}+R_{11.11}T_{11.11}\bar R_{11.11}
+R_{11.11}T_{11.00}\bar R_{00.11},\nonumber\\
\bar R_{00.11}(z)&=&R_{00.00}T_{00.11}\bar R_{11.11}
+R_{00.00}T_{00.00}\bar R_{00.11},\nonumber\\
\bar R_{11.00}(z)&=&R_{11.11}T_{11.11}\bar R_{11.00}
+R_{11.11}T_{11.00}\bar R_{00.00},\nonumber\\
\bar R_{00.00}(z)&=&R_{00.00}+R_{00.00}T_{00.11}\bar R_{11.00}
+R_{00.00}T_{00.00}\bar R_{00.00}.
\label{3.5}
\end{eqnarray}
For instance, the first one expresses the possibility of transition from (11) to itself either directly 
(without connecting vertices), or with at least one connecting vertex that involves $T_{11.11}$ 
or $T_{11.00}$.
The system (\ref{3.5}) can be solved easily, leading to  [cf. Ref. \cite{dH03b}, $\S4$]
\begin{eqnarray}
{\bar R}_{11.11}(z)&=&
\frac{z-W_{00.00}(z)}
{z\left(z-W_{11.11}(z)-W_{00.00}(z)\right)},
\label{3.6}
\\
{\bar R}_{00.00}(z)&=&
\frac{z-W_{11.11}(z)}
{z\left(z-W_{11.11}(z)-W_{00.00}(z)\right)},
\label{3.7}
\\
\bar R_{11.00}(z)&=& \frac{T_{11.00}(z)}{z({z-W_{11.11}(z)-W_{00.00}(z)})},
\label{3.8}
\\
\bar R_{00.11}(z)&=& \frac{T_{00.11}(z)}{z({z-W_{11.11}(z)-W_{00.00}(z)})}.
\label{3.9}
\end{eqnarray} 
The relevant analytical properties of those expressions have been studied in Ref. \cite{dH03b}.
The coincidence of the pole $z=\bar \theta$ 
with the corresponding pole at $z=\theta$
for the Liouvillian $L$ has been established.
The residues at those poles are well defined,
although the functions $W_{11.11}(z)$ and $W_{00.00}(z)$
present also branch points for the values of $z=0$ and 
$z=\bar \theta=\theta=\zeta+\bar \zeta$.
 
$\tilde{\bar \Sigma}_{aa.bb}(t)$ is then computed  from 
\begin{equation}
\tilde{\bar \Sigma}_{aa.bb}(t)
= \frac {-1}{2\pi i}\oint dz \, e^{-izt} 
\tilde{\bar R}_{aa.bb}(z)
=\frac {-1}{2\pi i}\oint dz \, e^{-izt}
\left(\frac 1{z-{\bar L}}\right)_{aa.bb}, 
\label{3.10}
\end{equation} 
 where the path contains only the two relevant physical poles:
The first one is the obvious $z=0$, the other one is a pole at $z=\bar \theta(=\theta=\zeta+\bar
\zeta)$.
The value $\bar \theta$ satisfies formally the equation
\begin{equation}
\bar \theta= W_{11.11}(\bar \theta)+W_{00.00}(\bar \theta),
\label{3.11a}
\end{equation} 
the solution does not depend of the chosen Riemann sheet.

We have by direct integration
\begin{equation}
\tilde{\bar \Sigma}_{11.11}(t)=
\frac{W_{00.00}(0)}
{\left(W_{11.11}(0)+W_{00.00}(0)\right)}
+ \exp{-i\bar \theta t}\,|\bar{{\cal A}}_1|^2 
\frac{W_{11.11}(\bar \theta)}
{\bar \theta}. 
\label{3.12}
\end{equation}
The residue at the pole $z=\theta$ has been noted as $|\bar{{\cal A}}_1|^2$.
That manner of writing has been introduced by analogy 
with the similar computation inside the rotating wave approximation \cite{dH03a}  
and does not imply a possible existence of $\bar{{\cal A}}_1$ 
for instance.
Indeed, we have shown in Ref. \cite{dH03b} the existence of $|\bar{{\cal A}}_1|^2$
as a sum of residues and not as a product of separate contributions.

The main difference with the treatment inside the RWA can be seen on that expression: 
two poles play a role in the expression of this (doubly) diagonal-diagonal element of 
$\tilde{\bar \Sigma}$.
Therefore, we have no longer the association of a temporal behaviour with a state.
Inside the RWA, the bare ground state coincides with the true ground state while we have now to
establish the connection.

Let us make some comments on the relative values of the contributions 
in a perturbation expansion in powers of the potential $V$.
Indeed, it is easily realized that $W_{00.00}(0)$ vanishes 
at the second order while $W_{11.11}(0)$ is finite at that order
and coincides with the second order value of $\bar \theta$.
This difference of behaviour has its origin in the stability 
of the state 0 at the lowest order.
Therefore, the second term starts as $1$ in a perturbation expansion in the potential 
while the first one is of order $V^2$.

The other elements of $\tilde{\bar \Sigma}$ are evaluated in appendix A.
The vacuum-vacuum elements of the super-operator $\tilde{\bar \Sigma}(t)$ 
for the value $t=0$ define the operator usually  called $\tilde{\bar A}$
in the subdynamics theory.
Its derivation and the obtention of its inverse is now straighforward.
The corresponding elements of the evolution operator $\tilde{\bar \Theta}$ 
in the vacuum subdynamics can be obtained from the relation
($P$ is the projector on the vacuum elements):
\begin{equation}
P\tilde{\bar \Sigma}(t)P=\exp{-i\tilde{\bar \Theta}t}\,\tilde{\bar  A}.
\label{3.28}
\end{equation}  
Therefore, by multiplying the time derivative of $\tilde{\bar \Sigma}(t)$
at $t=0$ by the inverse of the operator $A$, we get:
\begin{eqnarray}
\tilde{\bar \Theta}_{11.11}&=&\bar \theta \frac{W_{11.11}(0)}
{\left(W_{11.11}(0)+W_{00.00}(0)\right)},\nonumber\\ 
\tilde{\bar \Theta}_{00.00}&=&\bar \theta \frac{W_{00.00}(0)}
{\left(W_{11.11}(0)+W_{00.00}(0)\right)},\nonumber\\ 
\tilde{\bar \Theta}_{00.11}&=&-\bar \theta \frac{W_{11.11}(0)}
{\left(W_{11.11}(0)+W_{00.00}(0)\right)},\nonumber\\ 
\tilde{\bar \Theta}_{11.00}&=&-\bar \theta \frac{W_{00.00}(0)}
{\left(W_{11.11}(0)+W_{00.00}(0)\right)}.
\label{3.29}
\end{eqnarray}
Some obvious comments are a direct consequence of the form (\ref{3.29}) of the 
$\tilde{\bar \Theta}$ operator.
A first remarkable point is that this result does not obviouly depend on  a choice of a Riemann
sheet: that expression requires only the existence of the pole $\bar \theta$: the $z=0$ values for
the elements of the irredictible operator $W$ are obviously well defined.
That operator $\tilde{\bar \Theta}$ preserves the norm of $\tilde {\bar\rho}$, 
thanks the obvious relations 
$\tilde{\bar \Theta}_{11.11}=-\tilde{\bar \Theta}_{00.11}$ and
$\tilde{\bar \Theta}_{11.00}=-\tilde{\bar \Theta}_{00.00}$.
One of the eigenvalue of $\tilde{\bar \Theta}$ 
vanishes while the other one is the sum 
$\bar \theta$ of its diagonal elements.
A unitary transformation can
lead to a Jordan form, 
with the cancelation of the elements $\Phi_{11.00}$ and $\Phi_{00.00}$ of a new evolution operator
$\Phi$. 
The structure obtained inside the rotating wave approximation 
would then be rederived in an exact way.
This point will be considered (see $\S$5)  after the examination of the properties
of the other elements of $\tilde{\bar \Sigma}$ 
that do not involve field particles and describe atomic dipolar moments

\section{The atomic dipolar moment in the reduced formalism}

\setcounter{equation}{0}

We are interested in this section in the elements 
without field 
$\tilde{\bar \Sigma}_{10.10}$, $\tilde{\bar \Sigma}_{01.01}$,
$\tilde{\bar \Sigma}_{10.01}$, $\tilde{\bar \Sigma}_{01.10}$    
that are not dynamically connected with the previous ones. 
Those elements play a role in determining 
the properties of the dipolar moment of the atom.
As in the previous section, the extension of dynamics plays no role, the index ``tilde" can be
dropped and express the elements in the reduced formalism in terms of the elements in the
unreduced one's. 
Our starting expressions are therefore the formal solutions [formulae (\ref{2.8}-\ref{2.9})]
for the evolution of $\rho$ and $\bar\rho$,
where the couples $(ab)$ and $(cd)$ refer to the off diagonal discrete states 
(10) and (01). 
This can be easily symbolized by using a notation ($a\bar a$): $\bar a$ is one when $a$ is 0 and vice
versa. 
An element of comparison is provided by studying first $R(z)$.
That resolvent of $L$ that plays a role in (\ref{2.8}) can be written 
as a convolution of the corresponding resolvents of $H$ and $-H$.
Since the interaction involves 
the change of occupation numbers of field particles
by one unit for each absorption or emission process, the states $a$ and $c$ on one hand, 
$b$ and $d$ on the  other hand have to be identical.
Only diagonal elements of the resolvent of $H$ and $-H$ are thus to be considered
and we have
\begin{equation}
R_{a\bar a.a\bar a}(z)=\left(\frac 1{z-L}\right)_{a\bar a.a\bar a}
=\frac {-1}{2\pi i}\int_{0<\Im u<\Im z} du \left(\frac {1}{u-H} \right)_{aa}
\left(\frac {1}{z-u+H}\right)_{\bar a\bar a},
\label{4.1}
\end{equation}
involving  the functions $\eta$ and $\bar\eta$ given in (\ref{2.3a}-\ref{2.4a}).
The analysis of the convolution enables \cite{ELOP66,dH03b} to ascertain the presence
of an isolated pole to both diagonal elements of $R$.
In terms of the corresponding poles of the Green's functions associated
with the Hamiltonians $H$ and $-H$ (see $\S$2 or  Ref. \cite{dH03b}), 
they are given respectively by 
$z= \omega_1+\zeta-\omega_0+\bar\delta$
with a residue ${\cal A}_1 \bar{{\cal A}}_0 $ for $(a\bar a)=(10)$ and by 
$z=\omega_0+\delta-\omega_1+\bar\zeta$
with a residue ${\cal A}_0 \bar{{\cal A}}_1$ for the other case $(a\bar a)=(01)$.

The computation of the elements of $\tilde{\bar \Sigma}$ is detailed in appendix C.
The poles that play a role are still located at
 $z=\omega_1+\zeta-\omega_0+\bar \delta \equiv \delta_{10}$
and  $z=\omega_0+ \delta-\omega_1+\bar\zeta \equiv \delta_{01}$.
They now appear together as poles of each element.
The result can be symbolized as:
\begin{eqnarray}
\tilde{\bar \Sigma}_{a\bar a.a\bar a}(t)&=&e^{-i\delta_{10} t}\alpha_{a\bar a.a\bar a} 
+ e^{-i\delta_{01} t} \beta_{a\bar a.a\bar a},
\nonumber\\
\tilde{\bar \Sigma}_{\bar a a.a\bar a}(t)&=&e^{-i\delta_{10} t}\alpha_{\bar a a.a\bar a}
+ e^{-i\delta_{01} t} \beta_{\bar a a.a\bar a},
\label{4.13}
\end{eqnarray}
where the values of $\alpha$ and $\beta$ can be read on the equations (\ref{4.12}).
In a perturbation expansion, $\alpha_{10.10}$ and $\beta_{01.01}$ start as 1
while the other elements behave at least as $V^2$, since they
involve connecting vertices.

The corresponding elements of the evolution operator $\tilde{\bar \Theta}$ 
in the vacuum subdynamics can also be obtained from the relation (\ref{3.28})
through the previous procedure ($|\tilde{\bar A}_D|$ is defined in (\ref{4.17}))
\begin{eqnarray}
\tilde{\bar \Theta}_{a\bar a .a\bar a}&=&\frac1{|\tilde{\bar A}_D|}
(\delta_{10} \alpha_{a\bar a .a\bar a}+\delta_{01} \beta_{a\bar a .a\bar a})
(\alpha_{\bar a a.\bar a a}+ \beta_{\bar a a.\bar a a}),
\nonumber\\ 
\tilde{\bar \Theta}_{\bar a a.a\bar a}&=&-\frac1{|\tilde{\bar A}_D|}
(\delta_{10} \alpha_{\bar a a.a\bar a}+\delta_{01} \beta_{\bar a a.a\bar a})
(\alpha_{\bar a a.a\bar a}+ \beta_{\bar a a.a\bar a}).
\label{4.27}
\end{eqnarray}
In opposition with the case met in the RWA, the operator $\tilde{\bar \Theta}$ is not diagonal inside
the sectors defined by the atomic dipolar moments.

The value of $\left(e^{ -i\tilde{\bar \Theta}t}\right)_{a\bar a.b\bar b}$ can be obtained directly from
the knowledge of the elements of $\tilde{\bar \Sigma}$ and $\tilde{\bar A}^{-1}$:
\begin{eqnarray}
\left(e^{ -i\tilde{\bar \Theta}t}\right)_{a\bar a.b\bar b}&=&
\left(\tilde{\bar \Sigma} \tilde{\bar A}^{-1}\right)_{a\bar a.b\bar b}(t) \nonumber\\&=&
\left(e^{-i\delta_{10} t}\alpha_{a\bar a.10} 
+ e^{-i\delta_{01} t} \beta_{a\bar a.10}\right)
\left(\tilde{\bar A}^{-1}\right)_{10.b\bar b} \nonumber\\&+&
\left(e^{-i\delta_{10} t}\alpha_{a\bar a.01} 
+ e^{-i\delta_{01} t} \beta_{a\bar a.01}\right)
\left(\tilde{\bar A}^{-1}\right)_{01.b\bar b}.  
\label{5.31}
\end{eqnarray}
We know from the expression of $\tilde{\bar \Sigma}(t)$
that the two eigenvalues of that operator $\tilde{\bar \Theta}$
are $\delta_{10}$ and $\delta_{01}$ for the elements under consideration.
Therefore, an invertible transformation can provide a diagonal evolution operator 
that has these two values as the only non-vanishing elements. 

\section{ Physical representation for the purely atomic part of
$\tilde{\bar\rho}$}                                                                

\setcounter{equation}{0}

Denoting   $\tilde{\bar\rho}_{at}(t)$ the elements of $\bar\rho$ that do not involve photons, in the
absence of an incident field, we have, by definition of $\tilde{\bar\Theta}^{at} $, the set of
equations:
\begin{equation}
\frac{\partial}{\partial t}\tilde{\bar\rho}_{at}(t)=-i\tilde{\bar\Theta}^{at} \tilde{\bar\rho}_{at}(t),
\label{5.1}
\end{equation}
where the relevant elements of the operators $\tilde{\bar\Theta}^{at}$ 
are given in (\ref{3.29}) for the diagonal-diagonal elements 
and in (\ref{4.27}) for the off-diagonal ones.
Let us examine first the diagonal-diagonal elements.
The elements 
$\tilde{\bar\Theta}_{00.00}$ and $\tilde{\bar\Theta}_{11.00}$
would vanish if $W_{00.00}(0)=0$.
We may check explicitly  this property in a perturbation expansion.
The second order value $W_{00.00}^{(2)}(0)$ 
involves only contributions that are present 
in the unreduced formalism and vanishes accordingly. 
We have therefore to consider the fourth order value $W_{00.00}^{(4)}(0)$.
After some algebra, a non-vanishing contribution, that involves
the connecting vertices  $\bar  L'_V$ (\ref{2.5}) is obtained
(see Appendix B for details of computation). 
\begin{eqnarray}
W_{00.00}^{(4)}(0)&=&-2\pi i\sum_k\sum_{k'} |V_{1|0k}|^2 |V_{1|0k'}|^2 
\delta(\omega_1-\omega_0-\omega_k)\frac1{(\omega_1-\omega_0+\omega_{k'})^2}.
\nonumber\\&&
\label{5.2}
\end{eqnarray}
Since we have 
\begin{equation}
W_{11.11}^{(2)}(0)=-2\pi i\sum_k |V_{1|0k}|^2\delta(\omega_1-\omega_0-\omega_k)
=\bar{\theta}^{(2)}, 
\label{5.3}
\end{equation}
the first non-vanishing contribution to $\tilde{\bar\Theta}_{00.00}$ 
is of the fourth order
\begin{equation}
\tilde{\bar\Theta}_{00.00}^{(4)}=\bar{\theta}^{(2)}\sum_{k'}|V_{1|0k'}|^2
\frac1{(\omega_1-\omega_0+\omega_{k'})^2}. 
\label{5.4}
\end{equation}
The denominator is positive defined and no regularisation (``$i\epsilon$") appears.
Obviously, $\tilde{\bar\Theta}_{00.00}^{(4)}$ does not vanish.
As a consequence, $\tilde{\bar\rho}_{00}$ does not describe the atom in its ground state.
For comparison, $\tilde{\bar\Theta}_{11.11}$ starts at the second order and we have
$\tilde{\bar\Theta}_{11.11}^{(2)}={\bar\theta}^{(2)}$.
We know from the expressions of $\tilde{\bar \Sigma}$ 
that the operator $\tilde{\bar\Theta}$, relevant for the diagonal-diagonal elements has two
eigenvalues,  0 and $\bar\theta$.
If we take into account the conservation of the norm,
the operator $\tilde{\bar\Theta}$ can be related to an  operator $\Phi$
that has the structure met inside the RWA:
\begin{eqnarray}
\Phi_{11.11}&=&\bar\theta, \qquad
\Phi_{00.11}=-\bar\theta,\qquad
\Phi_{00.00}=\Phi_{11.00}=0. 
\label{5.6}
\end{eqnarray}
Let us note that the natural structure for the evolution operator
does not require an hermitian operator (or star-hermitian), in opposition to the early attempts in
the subdynamics approach.\cite{PGHR73}\ 
We define a new reduced density operator
$\tilde{\bar\rho}^P$ (in the so-called physical representation) connected to the original one
$\tilde{\bar\rho}$ through an invertible dressing operator $\chi$
\begin{equation}
\tilde{\bar\rho}^P(t)=\chi^{-1}\tilde{\bar\rho}(t) ,\qquad\qquad \tilde{\bar\rho}(t) =\chi
\tilde{\bar\rho}^P(t),
\label{5.7}
\end{equation}
and such that the evolution of $\tilde{\bar\rho}^P$ is governed by the operator $\Phi$.
Therefore, the following relations have to hold:
\begin{equation}
\Phi=\chi^{-1}\tilde{\bar\Theta}\chi, \qquad\qquad \tilde{\bar\Theta} =\chi \Phi \chi^{-1}.
\label{5.8}
\end{equation}
If we impose that the traces of $\tilde{\bar\rho}$ and $\tilde{\bar\rho}^P$ are the same,
we have moreover
\begin{eqnarray}
(\chi^{-1})_{11.11}+(\chi^{-1})_{00.11}&=&1, \qquad
(\chi^{-1})_{00.00}+(\chi^{-1})_{11.00}=1, \nonumber\\
\chi_{11.11}+\chi_{00.11}&=&1, \qquad
\chi_{00.00}+\chi_{11.00}=1. 
\label{5.9}
\end{eqnarray}
We have therefore to determine $\chi$ by imposing conditions (\ref{5.9})
and (\ref{5.8}).
The last conditions (\ref{5.8}) can be made explicit, taking (\ref{5.6}) into account:
\begin{equation}
\tilde{\bar\Theta}_{aa.bb}= \chi_{aa.11} \Phi_{11.11}(\chi^{-1})_{11.bb}
+ \chi_{aa.00} \Phi_{00.11}(\chi^{-1})_{11.bb}.
\label{5.11}
\end{equation}
The values of the elements of $\chi$ are computed from these conditions in Appendix D and
the unambiguous result is
\begin{eqnarray}
\chi_{11.11}=\chi_{00.00}&=& \frac{W_{11.11}(0)}{W_{11.11}(0)+W_{00.00}(0)}, \nonumber\\
\chi_{11.00}=\chi_{00.11}&=& \frac{W_{00.00}(0)}{W_{11.11}(0)+W_{00.00}(0)}, \nonumber\\
(\chi^{-1})_{11.11}=(\chi^{-1})_{00.00}&=&
\frac{W_{11.11}(0)}{W_{11.11}(0)-W_{00.00}(0)},\nonumber\\
(\chi^{-1})_{00.11}=(\chi^{-1})_{11.00}&=&
-\frac{W_{00.00}(0)}{W_{11.11}(0)-W_{00.00}(0)}.
\label{5.21}
\end{eqnarray}
The form of the operator $\Phi$ and the condition on the norm determines thus entirely the dressing
operator $\chi$ for these elements.
The two diagonal elements of $\tilde{\bar\rho}^P$ can thus be interpreted as the probability of
finding the atom resp. in the ground and excited levels (cf. \cite{dH03a}).
They are normalized and have the expected associated ``free" motion (in the absence of an incident
field).

We now turn to the elements describing the dipolar atomic momentum.
We know from the expressions of $\tilde{\bar \Sigma}$ 
that the operator $\tilde{\bar\Theta}$ has two complex eigenvalues
inside the considered subspace, namely  
$\delta_{10}$ and $\delta_{01}$.
The operator $\tilde{\bar\Theta}$ should be related to the operator $\Phi$,
the elements of which are given by:
\begin{equation}
\Phi_{10.10}=\delta_{10}, \qquad
\Phi_{01.01}=\delta_{01}, \qquad
\Phi_{01.10}=\Phi_{10.01}=0. 
\label{5.22}
\end{equation}
The natural structure for these elements evolution operator
requires here an hermitian operator, 
as opposed to the case of the diagonal elements.
The relevant elements of the reduced density operator $\tilde{\bar\rho}^P$
 are connected with the original one $\tilde{\bar\rho}$ through the same relation (\ref{5.7})
and its evolution is governed by the operator $\Phi$ (\ref{5.22}).
Therefore, the same relations (\ref{5.8}) have to hold inside the sector, but no condition on the
trace appears.

Denoting by $|\chi_D|$ 
the determinant of the $\chi$ matrix in the dipolar sectors,
we have from the inversion of a standard $2\times 2$ matrix
\begin{equation}
(\chi^{-1})_{a\bar a.a\bar a}=\chi_{\bar a a.\bar a a}\frac1{|\chi_D|},\qquad
(\chi^{-1})_{a\bar a.\bar a a}=-\chi_{a\bar a.\bar a a}\frac1{|\chi_D|}.
\label{5.25}
\end{equation}
We have therefore to determine $\chi$ by imposing conditions (\ref{5.8})
that can be made explicit, taking (\ref{5.22}) into account
\begin{equation}
\tilde{\bar \Theta}_{a\bar a.c\bar c}= 
\sum_b\chi_{a\bar a.b\bar b} \delta_{b\bar b}(\chi^{-1})_{b\bar b.c\bar c} .
\label{5.27}
\end{equation}
We know (\ref{4.27}) the explicit value of the elements of $\Theta$,
but we will not proceed by direct identification of (\ref{5.27}) and (\ref{4.27}).
In order to obtain the values af the elements of $\chi$,
it seems easier to compare the expressions of $\exp -i\tilde{\bar\Theta}t$ deduced from 
 $\chi (\exp -i\Phi t )\chi^{-1}$ and from its expression (\ref{5.31}).
We proceed in appendix D with the  identification of the terms with the same exponential behaviour
($\exp -i\delta_{10} t$ or $\exp -i\delta_{01} t$).
The elements of $\chi$ can be determined up to a free parameter.
A remaining indetermination is not new inside the context of subdynamics.\cite{PGHR73}\
It does not affect the evolution equations but concerns the relation between $\tilde{\bar\rho}^P$ and
$\tilde{\bar\rho}$.
That indetermination is linked to a choice of relative phase between the basic states used for
the description. 
The dressing operator is indeed apt to change the original choice.
We can fix it by  connecting, for instance, the value of ${|\chi|}$ with that of 
${|\tilde{\bar A}_D|}$ in a ``usual way":
${|\chi|^2}=|\tilde{\bar A}_D|$ or introduce a  criterion that ensures the positivity of the density
operator if required. 
 We will not elaborate here further on this point.

For the elements considered, 
the reduced density operator $\tilde{\bar\rho}^P$ satisfies
the kinetic equations that we have postulated in Ref. \cite{dH91}.
That derivation is exact (no approximation has been required).
The kinetic equations appear naturally when the physical representation
is introduced.
As already stressed in Ref. \cite{dH91},  physicists in optics use them
naturally, without realizing their profound origin, 
while  considering conceptually such equations 
as useful approximated equations arising from some pole approximation.
The formulation of observables in the formalism $\bar\rho^P$ is more
natural that in the original representation.
What we call the ground state and excited states of the atom are the 
objects that behave as the diagonal elements of $\bar\rho^P$:
in the absence of an incident field,
the excited state decays in a purely exponential way while 
the evolution of the ground state is only due to the transfer 
arising from the decay of the excited state.
Such a decay takes place with the lifetime as it is usually computed from 
$S$-matrix or Green's function formalism.
Our formalism, that eliminates the reabsorption of the field by its source,
leads automatically to a  kinetic description 
while preserving the equivalence with the original Liouville-von Neumann
equation, providing the compatibility conditions 
are satisfied at the initial time.\cite{dHG03}\
The atomic observables, computable with ${\bar\rho}_{at}$ ($=\tilde{\bar\rho}_{at}$ by the
constitutive relations),
have also to be modified by the dressing operator ($\chi$ or more
precisely
$\chi^{-1}$) if they are defined in the
original representation,  so that the conservation of their mean value is ensured.
If the observables are defined by the kinetic properties, such as the  probability of finding the atom
in the ground or excited state, they can be expressed directly in the physical representation.
Our approach enables moreover a consistent departure from the original time-reversible
description while preserving normalizability and positivity.\cite{dH03a}\
The element  $\bar\rho^P_{11}$ has a clear physical meaning as the probability of finding
the atom in the excited state, irrrespectively of the state of the field.
Inside RWA, in \cite{dH03a}, we have shown that an initial condition describing the atom in the
excited state (only $\bar\rho^P_{11}$ does not vanish) does not belong to an admissible initial
condition if the equivalence conditions are satisfied.
Such an initial condition is however fit to discuss normalizability and positivity in the irreversible
departure from quantum mechanics but its relevance in physics has to be established on an
experimental basis.

\section{One incident field}

\subsection{One passive incident field}

\setcounter{equation}{0}

We look first for the description of the elements of 
the evolution operators $\Theta$ and $\Phi$  
involving an incident field line 
that is not absorbed.
They can be obtained by considering the elements of $\tilde{\bar \Sigma}(t)$ 
that involve one incident field line at the extreme right and the same
field line at the left.
The atomic states are explicit in our notations 
and are represented by the letters $a$, $b$,$c$, $d$
($a$, $b$, $c$, $d$ represent the atomic state 0 or 1), while the present photons 
are written at the right of the atomic state.
We will consider therefore the following elements of $\tilde{\bar \Sigma}$:
$\tilde{\bar \Sigma}_{a\lambda b.c\lambda d}$, 
$\tilde{\bar \Sigma}_{ab\lambda .cd\lambda}$.
The couple of atomic states ($ab$) and ($cd$) involved 
at the right ($cd$) and at the left ($ab$)  of the elements of
$\tilde{\bar \Sigma}$ are of the same nature to get a non-vanishing contribution, namely
a diagonal couple or an off diagonal couple on both sides.

Our starting point is the formal expression of the contributions to 
$\bar\rho_{a\lambda b}(t)$ arising from the part of initial conditions 
involving only one incident field line:
\begin{equation} {\bar\rho}_{a\lambda b}(t) \leftarrow \sum_{c,d}\frac {-1}{2\pi i}
\int_c dz \, e^{-izt}
\left(\frac 1{z-\tilde{\bar L}}\right)_{a\lambda b.c\lambda d} {\bar\rho}_{c\lambda d}(0)
\label{6.1}
\end{equation}
The detailed calculation can be found in Appendix E and provides the expected result.
If we define $I_{ab.cd}$ as 1 when $a=c$ and $b=d$
($I_{ab.cd}$ vanishes for the other possibilities)
we obtain
\begin{equation}
\tilde{\bar\Theta}_{a\lambda b.c\lambda d}=\omega_\lambda I_{ab.cd}+
\tilde{\bar\Theta}_{ab.cd}\qquad
\tilde{\bar\Theta}_{ab\lambda .cd\lambda }=-\omega_\lambda I_{ab.cd}+
\tilde{\bar\Theta}_{ab.cd}
\label{6.15}
\end{equation}
The dressing does not depend on the incident photon and 
the natural choice for the dressing operator $\chi$  is 
\begin{equation}
\chi_{a\lambda b.c\lambda d}=\chi_{ab.cd}\qquad
\chi_{ab\lambda .cd\lambda }=\chi_{ab.cd}
\label{6.16}
\end{equation}
and therefore
\begin{equation}
\Phi_{a\lambda b.c\lambda d}=\omega_\lambda I_{ab.cd}+
\Phi_{ab.cd}\qquad
\Phi_{ab\lambda .cd\lambda }=-\omega_\lambda I_{ab.cd}+
\Phi_{ab.cd}
\label{6.17}
\end{equation}
The generalization of that property 
in the case of the presence of an arbitrary numbers of field lines
(incident or emitted) is obvious as long as they do not interact
with the atomic variables.

\subsection{ One absorbed incident field line}
We look for the elements of 
the evolution operators $\Theta$ and $\Phi$ that describe 
the absorption of one incident field line.
They can be obtained by considering the elements of $\tilde{\bar \Sigma}(t)$ 
that involve one incident field at the extreme right and
no field at the left.
We consider therefore the following elements of $\tilde{\bar \Sigma}$
($a$, $b$, $c$, $d$ represent the atomic states 0 or 1)
$\tilde{\bar \Sigma}_{ab.c\lambda d}$, $\tilde{\bar \Sigma}_{ab.cd\lambda}$.
The couple of atomic states ($ab$) and ($cd$) involved 
at the right ($cd$) and at the left ($ab$)  of
$\tilde{\bar \Sigma}$ are of different natures:
We have a diagonal couple on one side 
and an off-diagonal couple on the other side.
The intrinsic evolution of that kind of couples 
have been studied in previous sections.
If the diagonal couple is at the left, we have to consider:
$\tilde{\bar \Sigma}_{11.1\lambda0}$, $\tilde{\bar \Sigma}_{00.0\lambda1}$,
$\tilde{\bar \Sigma}_{11.0\lambda1}$, $\tilde{\bar \Sigma}_{00.1\lambda0}$,
$\tilde{\bar \Sigma}_{11.10\lambda}$, $\tilde{\bar \Sigma}_{00.01\lambda}$,
$\tilde{\bar \Sigma}_{11.01\lambda}$, $\tilde{\bar \Sigma}_{00.10\lambda}$.
Similar elements have to be considered in the case 
where the diagonal elements are at the right:
$\tilde{\bar \Sigma}_{10.1\lambda1}$, $\tilde{\bar \Sigma}_{10.0\lambda0}$,
$\tilde{\bar \Sigma}_{01.1\lambda1}$, $\tilde{\bar \Sigma}_{01.0\lambda0}$,
$\tilde{\bar \Sigma}_{10.11\lambda}$, $\tilde{\bar \Sigma}_{10.00\lambda}$,
$\tilde{\bar \Sigma}_{01.11\lambda}$, $\tilde{\bar \Sigma}_{01.00\lambda}$.

Our starting point is the formal expression of the contributions to 
$\bar\rho_{ab}(t)$ arising from the part of initial conditions 
involving only one incident field 
\begin{eqnarray} {\bar\rho}_{ab}(t) &\leftarrow&
\sum_{c,d,\lambda}\frac {-1}{2\pi i}\int_c dz \, e^{-izt}
\nonumber\\&&\times
\left[\left(\frac 1{z-\tilde{\bar L}}\right)_{ab.c\lambda d} {\bar\rho}_{c\lambda d}(0)
+\left(\frac 1{z-\tilde{\bar L}}\right)_{ab.cd\lambda } {\bar\rho}_{cd\lambda}(0) 
\right].
\label{7.1}
\end{eqnarray}
The details of the computation can be found in Appendix E and $\tilde{\bar\Theta}$ takes the form:
\begin{eqnarray}
\tilde{\bar\Theta}_{aa.b\lambda\bar b}&=&
\sum_{c,d,e} (\tilde{\bar\Theta} \tilde{\bar A})_{aa.cc} 
(\tilde{\bar A}^{-1})_{cc.dd}  \tilde{\bar A}_{dd.e\lambda \bar e}
(\tilde{\bar A}^{-1})_{e\bar e.b\bar b}+
\sum_c (\tilde{\bar\Theta} \tilde{\bar A})_{aa.c\lambda\bar c} 
(\tilde{\bar A}^{-1})_{c\bar c.b\bar b}\nonumber\\&=&
\sum_{d,e} \tilde{\bar\Theta}_{aa.dd} 
\tilde{\bar A}_{dd.e\lambda \bar e}
(\tilde{\bar A}^{-1})_{e\bar e.b\bar b}+
\sum_c (\tilde{\bar\Theta} \tilde{\bar A})_{aa.c\lambda\bar c} 
(\tilde{\bar A}^{-1})_{c\bar c.b\bar b}.
\label{7.32}
\end{eqnarray}
Similar expressions hold for the other elements $\tilde{\bar\Theta}_{aa.b\bar b\lambda}$,
$\tilde{\bar\Theta}_{a\bar a.b\lambda b}$, $\tilde{\bar\Theta}_{a\bar a.bb\lambda}$.

The derivation of the elements describing an emission process can be performed in a completely
similar way.
We will not dwell on these terms since they do not introduce new ideas or properties.

These expressions may serve to determine the corresponding elements of the dressing operator
$\chi$ and of the evolution  operator  $\Phi$ (see next $\S$7).

\section{Vertex dressing}

\setcounter{equation}{0}

We are interested in the possibilities for the evolution operator $\Phi$ that
describes the absorption of  one incident field line. 
We use our previous knowledge of the elements of $\chi$, that do not involve the field.
These elements have been determined in previous sections, using physical requirements.
For the computation of the element of $\Phi$ describing the disparition of the incident field line, 
we may use the link (\ref{5.8}) between $\tilde{\bar \Theta}$ and $\Phi$ and introduce a new element
$\chi_{gh.c\lambda d}$ for the dressing operator to get
\begin{eqnarray}
\Phi_{ab.c\lambda d}&=&\left(\chi^{-1}\tilde{\bar \Theta}\chi\right)_{ab.c\lambda d}
\nonumber\\&=&
\sum_{e,f,g,h} 
(\chi^{-1})_{ab.ef}
\tilde{\bar \Theta}_{ef.gh}
\chi_{gh.c\lambda d}
+\sum_{e,f,g,h} 
(\chi^{-1})_{ab.ef}
\tilde{\bar \Theta}_{ef.g\lambda h}
\chi_{gh.c\lambda d}
\nonumber\\&+&
\sum_{e,f,g,h} 
(\chi^{-1})_{ab.e\lambda f}
\tilde{\bar \Theta}_{e\lambda f.g\lambda h}
\chi_{g\lambda h.c\lambda d}.
\label{8.1}
\end{eqnarray} 
Let us call $\Phi^{at}$ the part of $\Phi$ that does not involve transitions in the field.
The elements of $\Phi^{at}$ have been determined in  $\S$5 [(\ref{5.6}) and (\ref{5.22})] and in  $\S$6
(\ref{6.17}).
Using this known value of $\Phi^{at}$, the property $\chi_{g\lambda h.c\lambda d}=\chi_{gh.cd}$
(see (\ref{6.16})) and the value of the element $(\chi^{-1})_{ab.e\lambda f}$ of the inverse of $\chi$,
that can be computed in the same manner as the corresponding element of the inverse of
$\tilde{\bar A}$ in appendix E, we get
\begin{eqnarray}
\Phi_{ab.c\lambda d}&=&
\sum_{e,f,g,h} 
(\chi^{-1})_{ab.ef}
\tilde{\bar \Theta}_{ef.g\lambda h}
\chi_{gh.c d}
+\sum_{e,f,g,h} 
\Phi^{at}_{ab.ef}
(\chi^{-1})_{ef.gh}
\chi_{gh.c\lambda d}
\nonumber\\&-&
\sum_{e,f,g,h} 
(\chi^{-1})_{ab.e f}
\chi_{ef.g\lambda h}
\Phi^{at}_{ g\lambda h.c\lambda d}.
\label{8.2}
\end{eqnarray} 
Defining the $X$ operator such that its only non-vanishing elements are
\begin{eqnarray}
X_{ab.c\lambda d}&=&
\sum_{e,f} 
(\chi^{-1})_{ab.e f}
\chi_{ef.c\lambda d},
\nonumber\\
X_{ab.cd\lambda }&=&
\sum_{e,f} 
(\chi^{-1})_{ab.e f}
\chi_{ef.cd\lambda},
\label{8.3}
\end{eqnarray} 
we can express the last  two terms in (\ref{8.2}) as a commutator of $X$ with the operator
$\Phi^{at}$.
\begin{eqnarray}
\Phi_{ab.c\lambda d}&=&
\sum_{e,f,g,h} 
(\chi^{-1})_{ab.ef}
\tilde{\bar \Theta}_{ef.g\lambda h}
\chi_{gh.c d}
+([\Phi^{at},X])_{ab.c\lambda d}
\nonumber\\ 
\Phi_{ab.cd\lambda}&=&
\sum_{e,f,g,h} 
(\chi^{-1})_{ab.ef}
\tilde{\bar \Theta}_{ef.gh\lambda }
\chi_{gh.c d}
+([\Phi^{at},X])_{ab.cd\lambda }.
\label{8.4}
\end{eqnarray} 
The elements $\Phi_{ab.c\lambda b}$ and $\Phi_{ab.cd\lambda }$ are expressed as the corresponding
elements of $\tilde{\bar \Theta}$, with a dressing bearing on the atomic levels,  plus the
commutator of $\Phi^{at}$ with an undetermined $X$ operator.
The role of that indetermination can be understood from the introduction of the dressed  reduced
density operator $\tilde{\bar \rho}^P$ (\ref{5.7}).
Indeed, the determination of the elements of $X$ is equivalent to that of $\chi$ [from (\ref{8.3})].
The presence of a non-vanishing $X$ operator means that the elements of the physical reduced
density operator  $\tilde{\bar\rho}^P_{ab}$ are also dressed by the elements 
$\tilde{\bar\rho}_{a\lambda b}$ and $\tilde{\bar\rho}_{ab\lambda}$: 
The dressing of the atomic states involves the incident photons.
This is by no means mandatory: we can choose to dress the states by the self-field only.
Therefore, the future determination of the $X$ operator requires a personal choice of the basic
states for describing the atom.\cite{CT94}\
Such a choice cannot be implied by the formalism.

Due to the structure of the commutator, the contribution to $\Phi$ depending on $X$ vanishes for the
resonnance processes when the width of the states is neglected:
This contribution plays a role in the off resonnance processes.
This point can be illustrated by considering some first order non-vanishing contributions to $\Phi$, 
the element $\Phi^{(1)}_{11.0\lambda 1}$ for instance.
We can replace in (\ref{8.4}) the $\chi$ and $\chi^{-1}$ operator by unity.
The first order contribution to $\tilde{\bar \Theta}_{11.0\lambda 1}$ is nothing but $V_{1|0\lambda}$.
$\Phi_{11.11}$ is at least from the second order $(\tilde{\theta}^{(2)})$ while 
$\Phi_{0\lambda 1.0\lambda 1}$ provides at the lowest order a contribution independent of the
coupling ($\Phi^{(0)}_{0\lambda 1.0\lambda 1}=\omega_0+\omega_\lambda-\omega_1$).
Therefore,
\begin{equation}
\Phi^{(1)}_{11.0\lambda 1}=
V_{1|0\lambda}-(\omega_0+\omega_\lambda-\omega_1)
X_{11.0\lambda 1}.
\label{8.5}
\end{equation}
If $X_{11.0\lambda 1}$ is not singular for $\omega_\lambda=\omega_1-\omega_0$, its value is
irrelevant for determining $\Phi^{(1)}_{11.0\lambda 1}$ at the bare energy resonnance
($\omega_\lambda=\omega_1-\omega_0$).

It has been long noticed  \cite{GH74}-\cite{GH98} that the kind of coupling 
(\ref{2.1}) between atoms and field is not completely satisfactory with respect 
to the causal propagation of the field, i.e. its  finite propagation. 
Indeed, in the problem of transfer of excitation from one atom to another, 
if the measurement concerns also the  state of the initially excited atom
(in a case of non local observables, as treated in case I of Ref. \cite{PT97})
precursors appear and  causality is not strictly respected.\cite{PT97}\
The same kind of acausal  behaviour is met \cite{dH85}
in the description of a double photodetection of the light from an atom 
admitting a cascade decay\footnote{In Ref. \cite{MJF95}, causality is claimed to be respected in
the case of a double photodetection but that property is limited to two-level systems while 
multilevel systems were considered in Ref. \cite{dH85}}.  
Those non causal behaviours are linked to
the  finite lower bound in the spectrum of the exchanged photon and   disappear when the integration
over the  energy of the photons is extended to
$-\infty$.  This was common practice. 
Various authors  \cite{SK30}-\cite{MK74} have claimed to have succeeded in proving causality of
propagation of light  : They all used  at some stage  an equivalent procedure. 
In the case of \cite{MK74}, for instance, the sign of a contribution is changed on the base of its
smallness.\cite{dH85}\
In later works, that include the counter-rotating terms, that procedure  has no longer been required
for proving strict causality in either the framework of two-level systems for all
observables,\cite{BCPPP90,CPPP95,MJF95}\ or, for multilevel systems for local
observables.\cite{PT97} \
Nevertheless, the general  case, beyond the two-level systems and local observables,
still contains acausal behaviours, even when the counter-rotating contributions are
considered.\cite{dH85,PT97}\

In Ref. \cite{dH85}, a suggestion has been proposed to ensure a strict causality in all possible cases
by proposing new terms of interaction between the atom and the field.
The present formalism is more appropriate for realizing that possibility of
including  in a simple way causality into the equations of motion 
without modifying the energy spectrum of the field. 
With respect to the usual approach, we have indeed a supplementary degree of freedom since the
symmetry beween absorption and emission is lost.\cite{dHG03,dH03a}\ 
In the study of the coupling of the quantum field with a quasiclassical source,\cite{dH91}  it has
been established  that an appropriate association of super-operators for the creation of the
field [${\cal E}_c$ in Ref. \cite{dH91}] and absorption process [$\cal E$ in Ref. \cite{dH91}] leads
to a strictly causal propagation, without precursor : 
The retarded solution for the field propagation
appears automatically, irrespectively of the atomic state.  
We can use the possibility of dressing to meet the causality
requirements. 
As we have seen, the dressing bears on  non-resonant contributions.
Therefore, we can require that the $\Phi$ elements, describing the interaction with an atom at some
point $\bf r$, have the appropriate structure for a local coupling ($\bf k$ is the wave number
associated with
$\lambda$ or $\mu$)
\begin{eqnarray}
\Phi_{ab.c\lambda d}&=&\phi^{abs}_{ab.cd}\frac1{\sqrt\omega_k}e^{i\bf k.\bf r}
,\qquad
\Phi_{ab.cd\lambda}=\phi^{abs}_{ab.cd}\frac1{\sqrt\omega_k}e^{-i\bf k.\bf r}
,\nonumber\\ 
\Phi_{a\mu b.cd}&=&\phi^{em}_{ab.cd}\frac1{\sqrt\omega_k}e^{-i\bf k.\bf r}
,\qquad
\Phi_{ab\mu.cd}=-\phi^{em}_{ab.cd}\frac1{\sqrt\omega_k}e^{i\bf k.\bf r}. 
\label{12.37}
\end{eqnarray}
The dressing by $\chi$ cannot change the value of $\Theta$
for the resonant contribution at the lowest order in the coupling but can provide any chosen value
for the other (off-resonant) elements.
It is thus perfectly possible to choose  $\chi$ such that these conditions are fulfilled.
The change of relative sign beween the two tems describing absorption and emission is capital.
When an emission, by a pointlike atom at some point ${\bf r}$, is combined with the absorption by
another one, at some other point ${\bf r}_1$, two processes lead to the same change in the atomic
occupation  numbers (\ref{12.37}). 
The field emitted by an atom through $\Phi_{a\mu b.cd}$, and absorbed by the other
atom through $\Phi_{a'b'.c'\lambda d'}$, provides a time dependence as $\exp (-i\omega_kt)$ while
the other process ($\Phi_{ab\mu.cd}$ and $\Phi_{a'b'.c'd'\lambda}$) involves $\exp (i\omega_kt)$.
Both contributions can be combined and their sum is
equivalent of having a domain of integration over 
the photon spectrum from $-\infty$ to $+\infty$, but this procedure is now an integral part of the 
evolution equation.
Therefore, the resulting contribution ignores the presence of a finite lower bound in the spectrum
of the field and causality can be fully respected, without precursors or other oddities:\cite{dH85}\
The propagator corresponding to a retarded solution for an electromagnetic process appears
naturally, corresponding to Ritz's point of view on the origin of irreversibility, rather than
Einstein's conception.\cite{RE09}\
 Of course, such a radical change in the form of the coupling has to take place while preserving
positivity of the retrieved density operator.

The emphasis on the physical interpretation of the elements of the description is not the only
difference with the Brussels-Austin group.\cite{OPP01,POP01}\
Indeed, as can be seen on the Friedrichs model,\cite{POP01} their approach is based on the
consideration of Gamow vectors obtained through a generalized eigenvalues problem, outside
Hilbert space, providing complex values. 
That obtention have been inspired by the early attempts of constructing a
subdynamics.\cite{dHH73a}\
 Besides problems linked to normalization (such states are of
null norm and a set of left and right bicomplete and biorthogonal  eigenvectors for the Hamiltonian
has to be introduced),  and positivity (that is not a requirement of their $\Lambda$
transformation), such states can only decay (by construction) and the description of an excitation of
the atom by an incident wave packet is outside the possibility of that approach.\footnote{That
criticism can also be addressed to our previous work \cite{dHH73a}  and has been one of our
motivations to depart from the early attemps of the Brussels group.}    
In contrast, the presence, in our kinetic operator $\Phi$, of non-diagonal elements in the field (see
(\ref{12.37}) for instance) enables a coupling between an incident wave packet and the atom.
Therefore, the work of that Brussels-Austin group is more appropriate for an abtract discussion
about irreversibility,\cite{RB04} through the introduction of ``diagonal singularities", than for the
description of atom-field interactions required in quantum optics.

\section{Concluding remarks} 

This paper enables to understand the link between reversible and irreversible formulations of
interacting atoms and fields.
This later formulation requires a description in a Schr\"odinger type description.
The existence to an invertible transformation enabling a transition from one description to the other
one may be perceived as a surprise: when compatibility conditions are satisfied, both formulations
are equivalent.
The physical representation enables physicists to provide in quantum optics a description conform
to their intuition, without any loss of generality or the introduction of approximations.
An atomic level is characterized by the values of its (dressed) energy and its
lifetime. 
Dressing contributions are now excluded through in the structure of the evolution operator
that  involves the often used distinction between external and emitted photons.
Quasiclassical state for the external field  can  be considered.\cite{dH91}\
The external field  description  naturally factorizes inside the reduced density operator.
Acausal behaviours can now be excluded through the choice of the dressing operator.
A modelization for an ideal  photodetection device \cite{dH85} finds naturally his place in that
framework. 
The relation with the original description involves the dressing operator, the
constitutive relations and the compatibility conditions.
A direct formulation of an initial condition for a problem inside the physical representation is
nevertheless possible, with the possibility of dealing with an extension of quantum mechanics, as has
been considered in Ref. \cite{dH03a}.
An initial condition where only the physical excited state is present cannot be obtained from an
initial condition in the original formulation.\cite{dHG03,dH03a}\
The necessity of such an extension of quantum mechanics cannot be excluded {\it a
priori} but should result from experiments.

This paper has focused on the role of dressing inside the single subdynamics approach.
In the first papers within that approach, no need for a dressing has appeared.
The main difference of the system inside and outside the RWA
is indeed the following: In presence of counter-rotating terms, 
the obtention of the ground and excited states of the atom
is no longer automatic and requires a dressing operator.
Indeed, inside the RWA, the bare ground state can be identified with the true ground state but that
property is lost outside the RWA:  It will not convert itself into an invariant state
and an adequate combination is required 
that enables the identification of the ground state and of the excited state
through their temporal dependences.
Renormalization through a dressing operator takes naturally place for dealing with physical
ground and excited states,  physical atomic dipolar moments,  pointlike interactions and 
causality. 
The recourse to these entities to make physical predictions depends of course of the
observables considered.
If they are defined in the original representation, no need to go into the physical representation is
present, except for the simplicity and transparency of the evolution in that representation.
If the observables are to be precized, they are more naturally defined inside the new representation.
Indeed, the precise meanings of the ground state, the excited state, the atomic dipolar moments
resort to $\tilde{\bar \rho}^P$ where temporal behaviours enable the identification.
The recourse to the new description can also be justified according to the manner in which the
initial conditions have to be formulated.
If we intend to consider the scattering of a wave packet impinging on an atom in its ground state,
we need to know how to describe correctly the atomic state and the present approach answers
that question.

We have not considered in this paper the constitutive and compatibility conditions since they
appear here in a fashion similar to our previous papers.
We are aware that positivity should require a more detailed analysis to place restrictions on the
undetermined elements of the dressing operator.

This  paper shows morever how the modelization in optics is justified 
from first principles without the need of approximations such as the pole approximation.
In our kinetic equations, the explicit attribution 
of a lifetime to an atomic excited level
is the consequence of the original Liouville-von Neumann equations
when going into an ``historical" representation.

The kinetic equations obtained in the present paper
are at the level of reduced distribution functions 
for the field, and have been introduced in \cite{dH91}.
The kinetic equations that were surmised in that paper are now properly
derived.

The dressing procedure introduced in the early approach
of the subdynamics theory has proved to be still relevant and fruitful
in the present work for more elaborate systems,
closer to quantum field theory.
We are nevertheless still facing some basic indetermination 
corresponding to the arbitrariness in a choice of basis vectors.
That indetermination is not met for the purely atomic part,
except for a (trivial) choice of the relative phase 
between the ground and excited atomic states: The atomic dipolar moment is not completely fixed.
The indetermination can modify the form of the evolution  operator
when absorption and emission processes are considered and can be used to ensure a strict causality,
in spite of a finite lower bound in the energy spectrum.

We have not dealt with simultaneous processes \cite{dH03a} that
are to be present both inside and outside the RWA,
the simultaneous absorption and emission process appears at a lower order 
outside the RWA than inside.
Since they involve necessarily nonresonant contributions,
they could be of the same importance 
as  higher order resonant contributions, not included into the model Hamiltonian.

\section*{Acknowledgements}
We  are pleased to thank hartily Professor C. George for numerous discussions and his help in
preparing this paper.

\appendix
\section{Other elements of $\tilde{\bar \Sigma}$ between diagonal atomic states}
\setcounter{equation}{0}

\def\theequation{
A\arabic{equation}}

We now turn to the second element $\tilde{\bar \Sigma}_{00.00}$.
From the following relations obtained in Ref. \cite{dH03b}, 
\begin{equation}
\bar R_{11.11}(z)+\bar R_{00.00}(z)=\frac1z+
\frac1{z-W_{11.11}(z)-W_{00.00}(z)},
\label{3.13}
\end{equation}  
we have
\begin{equation}
\tilde{\bar \Sigma}_{11.11}(t)+\tilde{\bar \Sigma}_{00.00}(t)=1+
\exp{-i\bar \theta t}\,|\bar{{\cal A}}_1|^2,
\label{3.14}
\end{equation}  
and therefore
\begin{equation}
\tilde{\bar \Sigma}_{00.00}(t)=
\frac{W_{11.11}(0)}
{\left(W_{11.11}(0)+W_{00.00}(0)\right)}
+ \exp{-i\bar \theta t}\,|\bar{{\cal A}}_1|^2 
\frac{W_{00.00}(\bar \theta)}
{\bar \theta}. 
\label{3.15}
\end{equation}
The first term starts as $1$ in a perturbation expansion in the potential
while the second one is of order $V^2$.

We now turn to the next element $\tilde{\bar \Sigma}_{11.00}$,
computed from (\ref{3.8}).
The contributions to $\tilde{\bar \Sigma}_{11.00}(t)$ emerge also from the
poles at $z=0$ and $z=\bar \theta$.
The existence of those poles, established in Ref. \cite{dH03b} for
$\tilde{\bar \Sigma}_{11.11}(t)$ and  $\tilde{\bar \Sigma}_{00.00}(t)$  
enables to prove their existence for $\tilde{\bar \Sigma}_{11.00}(t)$.
We can easily see that the residues of those poles are well defined,
in particular that the numerator is uniquely defined for that value.
Indeed, by comparing thir perturbation expansion 
(the relevant part of the unpertubed propagator in reduced formalism  
$\bar R^0(z)$ coincides with $R^0(z)$), 
we have
\begin{eqnarray}
&&\bar R_{00.00}(z)=
\sum_{n=0}^{\infty} \left(R^0(z) 
\left[\bar L_V R^0(z)\right]^n\right)_{00.00},\nonumber\\
&&\bar R_{11.00}(z)=
\sum_{n=0}^{\infty} \left(R^0(z) 
\left[\bar L_V R^0(z)\right]^n\right)_{11.00}.
\label{3.16}
\end{eqnarray}
If in the second expression (of $\bar R_{11.00}(z)$), 
we modify the last vertex we get a contribution
to $\bar R_{00.00}(z)$ with a change of sign.
The term $n=0$, without vertex $\bar L_V$, of $\bar R_{00.00}(z)$ cannot be
recovered in that way and we have to add its contribution.
Similar considerations hold als when the last two indices ar ``11" in place of ``00" and we have
\begin{eqnarray}
&&\bar R_{11.00}(z)= -\bar R_{00.00}(z)+\frac1z
\label{3.17}
\\
&&\bar R_{00.11}(z)= -\bar R_{11.11}(z)+\frac1z,
\label{3.22}
\end{eqnarray}
We can check the compatibility of the two expressions (\ref{3.8}) and (\ref{3.17}).
Using the expression (\ref{3.7}) for $\bar R_{00.00}(z)$ we get:
\begin{eqnarray}
\bar R_{11.00}(z)&=&-\bar R_{00.00}(z)+\frac1z =
-\frac{z-W_{11.11}(z)} 
{z\left(z-W_{11.11}(z)-W_{00.00}(z)\right)}
+\frac1z \nonumber\\
&=&\frac{-W_{00.00}(z)} 
{z\left(z-W_{11.11}(z)-W_{00.00}(z)\right)}. 
\label{3.18}
\end{eqnarray}
If we take into account the relation (\ref{3.4})
between $W_{00.00}(z)$ and $T_{11.00}(z)$,
we get the equivalence between the two forms (\ref{3.8}) and (\ref{3.17})
for $\bar R_{11.00}(z)$.
Its interest lies in the following remark:
in Ref. \cite{dH03b}, only the existence of the poles 
for the elements $\bar R_{11.11}(z)$ and $\bar R_{00.00}(z)$
has been considered.
The formulae (\ref{3.17}), (\ref{3.22}) show directly that the poles of $\bar R_{11.00}(z)$ and $\bar
R_{00.11}(z)$ are automatically well defined.
We get therefore:
\begin{equation}
\tilde{\bar \Sigma}_{11.00}(t)=
 \frac{-T_{11.00}(0)  }{\left(W_{11.11}(0)+W_{00.00}(0)\right)}   
+\exp{-i\bar \theta t}\, |\bar{{\cal A}}_1|^2
\frac{T_{11.00}(\bar \theta)}{\bar \theta},
\label{3.19}
\end{equation}
and similarly:
\begin{equation}
\tilde{\bar \Sigma}_{00.11}(t)=
 \frac{-T_{00.11}(0)  }{\left(W_{11.11}(0)+W_{00.00}(0)\right)} 
+\exp{-i\bar \theta t}\, |\bar{{\cal A}}_1|^2
T_{00.11}(\bar \theta).
\label{3.20}
\end{equation}
An alternative expression for $\tilde{\bar \Sigma}_{11.00}(t)$ and $\tilde{\bar \Sigma}_{00.11}(t)$
can be obtained from (\ref{3.17}), (\ref{3.22})  or from (\ref{3.3})
\begin{equation}
\tilde{\bar \Sigma}_{11.00}(t)=
\frac{W_{00.00}(0)}
{\left(W_{11.11}(0)+W_{00.00}(0)\right)}
- \exp{-i\bar \theta t}\,|\bar{{\cal A}}_1|^2
\frac{W_{00.00}(\bar \theta)}
{\bar \theta} 
\label{3.21}
\end{equation}
\begin{equation}
\tilde{\bar \Sigma}_{00.11}(t)=
\frac{W_{11.11}(0)}
{\left(W_{11.11}(0)+W_{00.00}(0)\right)}
- \exp{-i\bar \theta t}\,|\bar{{\cal A}}_1|^2 
\frac{W_{11.11}(\bar \theta)}
{\bar \theta}. 
\label{3.23}
\end{equation}
Let us note the obvious relations, 
that can be derived from (\ref{3.17}) and (\ref{3.22})
\begin{equation}
\tilde{\bar \Sigma}_{00.11}(t)+\tilde{\bar \Sigma}_{11.11}(t)= 1,\qquad
\tilde{\bar \Sigma}_{00.00}(t)+\tilde{\bar \Sigma}_{11.00}(t)= 1.
\label{3.24}
\end{equation}
Those relations  (\ref{3.12}), (\ref{3.15}), (\ref{3.21}) and (\ref{3.23}) can be symbolized as:
\begin{eqnarray}
\tilde{\bar \Sigma}_{aa.bb}(t)&=&\alpha_{aa.bb} 
+  e^{-i\theta t}\beta_{aa.bb},
\label{4.13a}
\end{eqnarray}
where the values of $\alpha$ and $\beta$ can be read on the preceding equations.
In a perturbation expansion, $\alpha_{11.11}$ and $\beta_{00.00}$ starts as 1
while the other elements behaves at least as $V^2$.

The vacuum-vacuum elements of the super-operator $\tilde{\bar \Sigma}(t)$ 
for the value $t=0$ define the operator usually  called $\tilde{\bar A}$
in the subdynamics theory.
The corresponding elements are
\begin{eqnarray}
\tilde{\bar A}_{aa.bb}(t)&=&\alpha_{aa.bb} + \beta_{aa.bb},
\label{4.13b}
\end{eqnarray}
or more explicitly
\begin{eqnarray}
\tilde{\bar A}_{11.11}&=&
\frac{W_{00.00}(0)}
{\left(W_{11.11}(0)+W_{00.00}(0)\right)}
+ |\bar{{\cal A}}_1|^2
\frac{W_{11.11}(\bar \theta)}
{\bar \theta} ,\nonumber\\
\tilde{\bar A}_{00.00}&=& \frac{W_{11.11}(0)}
{\left(W_{11.11}(0)+W_{00.00}(0)\right)}
+|\bar{{\cal A}}_1|^2 
\frac{W_{00.00}(\bar \theta)}
{\bar \theta} \nonumber\\
\tilde{\bar A}_{11.00}&=&
\frac{W_{00.00}(0)}
{\left(W_{11.11}(0)+W_{00.00}(0)\right)}
- |\bar{{\cal A}}_1|^2 
\frac{W_{00.00}(\bar \theta)}
{\bar \theta},\nonumber\\
\tilde{\bar A}_{00.11}&=&
\frac{W_{11.11}(0)}
{\left(W_{11.11}(0)+W_{00.00}(0)\right)}
-|\bar{{\cal A}}_1|^2 
\frac{W_{11.11}(\bar \theta)}
{\bar \theta}. 
\label{3.25}
\end{eqnarray}
The corresponding elements of the inverse $\tilde{\bar A}^{-1}$
of the $\tilde{\bar A}$ operator
can already be computed, independently of the elements of the subdynamics superoperator
involving the  field 
by the inversion  of a two by two matrix.
The determinant ${\cal A}_D$ of that matrix is
\begin{equation}
{\cal A}_D=\tilde{\bar A}_{11.11} \tilde{\bar A}_{00.00}
-\tilde{\bar A}_{00.11}\tilde{\bar A}_{11.00}= 
|\bar{{\cal A}}_1|^2,
\label{3.26}
\end{equation}  
as can be shown by direct computation using (\ref{3.11a}).
We have therefore directly
\begin{eqnarray}
\left(\tilde{\bar A}^{-1}\right)_{11.11}&=&
\frac{W_{11.11}(0)}
{\left(W_{11.11}(0)+W_{00.00}(0)\right)} 
|\bar{{\cal A}}_1|^{-2} 
+ 
\frac{W_{00.00}(\bar \theta)}
{\bar \theta}, 
\nonumber\\
\left(\tilde{\bar A}^{-1}\right)_{00.00}&=&  
\frac{W_{11.11}(0)}
{\left(W_{11.11}(0)+W_{00.00}(0)\right)}
|\bar{{\cal A}}_1|^{-2}
+ \frac{W_{00.00}(\bar \theta)}
{\bar \theta},  
\nonumber\\
\left(\tilde{\bar A}^{-1}\right)_{11.00}&=&
-\frac{W_{00.00}(0)}
{\left(W_{11.11}(0)+W_{00.00}(0)\right)}
|\bar{{\cal A}}_1|^{-2}
+\frac{W_{00.00}(\bar \theta)}
{\bar \theta},
\nonumber\\
\left(\tilde{\bar A}^{-1}\right)_{00.11}&=&
-\frac{W_{11.11}(0)}
{\left(W_{11.11}(0)+W_{00.00}(0)\right)}
|\bar{{\cal A}}_1|^{-2}
+\frac{W_{11.11}(\bar \theta)}
{\bar \theta}. 
\label{3.27}
\end{eqnarray}

\section{The fourth order contribution to $W_{00.00}(0)$ }  
\setcounter{equation}{0}

\def\theequation{
B.\arabic{equation}}

We derive in this appendix the expression (\ref{5.2}) 
for $W_{00.00}^{(4)}(0)$.
We have to consider all irreductible contributions leading 
from the diagonal state $00$ to itself.
We have 16 contributions to evaluate.
Our convention to denote a matrix element 
$<a|\bar \rho|b>=\bar \rho_{ab}$ of the (reduced) density operator $\bar \rho$
is to write for $a$ and $b$ in the first place the state of the atom.
This convention avoids the need of a separator when writing the indices $ab$.
Using that convention, the possible succession of correlated states
to be considered to evaluate $W_{00.00}$ from (\ref{4.14}) 
can be described in the following way.

We first  consider contributions that do not involve 
the connecting vertex $\bar  L'_V$ and are included in $\psi^{(4)}_{00.00}$.
They correspond to the following successions
that involve as last vertex (extreme left)
a change of the first index:
\newline
\noindent
(00;1k0;0kk'0;1k0;00), (00;1k0;0kk'0;1k'0;00),
(00;1k0;1k1k';1k0;00), 
\newline
\noindent
(00;1k0;1k1k';01k';00).
The corresponding contributions are called 
$C_1(z)$, $C_2(z)$, $C_3(z)$, $C_4(z)$.
They are:
\begin{eqnarray}
C_1(z)&=&\sum_k\sum_{k'} |V_{1|0k}|^2 |V_{1|0k'}|^2
\frac1{z-\omega_1+\omega_0-\omega_k}\frac1{z-\omega_k-\omega_{k'}}
\nonumber\\&\times&
\frac1{z-\omega_1+\omega_0-\omega_k} ,\nonumber\\
C_2(z)&=&\sum_k\sum_{k'} |V_{1|0k}|^2 |V_{1|0k'}|^2
\frac1{z-\omega_1+\omega_0-\omega_k}\frac1{z-\omega_k-\omega_{k'}}
\nonumber\\&\times&
\frac1{z-\omega_1+\omega_0-\omega_{k'}} ,\nonumber\\ 
C_3(z)&=&\sum_k\sum_{k'} |V_{1|0k}|^2 |V_{1|0k'}|^2
\frac1{z-\omega_1+\omega_0-\omega_k}\frac1{z-\omega_k+\omega_{k'}}
\nonumber\\&\times&
\frac1{z-\omega_1+\omega_0-\omega_k} ,\nonumber\\ 
C_4(z)&=&\sum_k\sum_{k'} |V_{1|0k}|^2 |V_{1|0k'}|^2
\frac1{z-\omega_1+\omega_0-\omega_k}\frac1{z-\omega_k+\omega_{k'}}
\nonumber\\&\times&
\frac1{z+\omega_1-\omega_0+\omega_{k'}}.    
\nonumber\\ && 
\label{A1.1}
\end{eqnarray}
The contributions involving as last vertex
a change of the second index (such as (00;01k;00kk';01k;00)) can be obtained form these expressions 
by replacing in the propagators $\omega \to -\omega$ 
for all values of the indices.
The corresponding contributions are called 
$C_{1'}$, $C_{2'}$, $C_{3'}$, $C_{4'}$.
The limit $z\to 0$ has to be considered to get the contributions to
$W_{00.00}^{(4)}(0)$.
For the terms $C_1$ and $C_2$, the limit $z\to 0$ 
can be taken in a harmless way
since no propagator can be resonant.
Therefore, we have
\begin{eqnarray}
C_1(0)&=&\sum_k\sum_{k'} |V_{1|0k}|^2 |V_{1|0k'}|^2
\frac1{-\omega_1+\omega_0-\omega_k}\frac1{-\omega_k-\omega_{k'}}
\frac1{-\omega_1+\omega_0-\omega_k} ,\nonumber\\
C_2(0)&=&\sum_k\sum_{k'} |V_{1|0k}|^2 |V_{1|0k'}|^2
\frac1{-\omega_1+\omega_0-\omega_k}\frac1{-\omega_k-\omega_{k'}}
\frac1{-\omega_1+\omega_0-\omega_{k'}}.    
\nonumber\\ && 
\label{A1.2}
\end{eqnarray}
For obvious reasons, we have therefore:
\begin{eqnarray}
C_1(0)+C_{1'}(0)&=&0 \qquad\qquad
C_2(0)+C_{2'}(0)=0   
\label{A1.3}
\end{eqnarray}
We now turn to the contributions $C_3$ and $C_4$.
The limit $z\to 0$ has to be taken carefully for the propagator
$(z-\omega_k+\omega_{k'})^{-1}$ since it is resonant for 
$\omega_k=\omega_{k'}$.
Let us consider the sum $C_{34}$ of $C_3$ and $C_4$ and take the limit  $z\to 0$ 
whenever it is harmless:
\begin{eqnarray}
C_{34}(0)&=&\lim_{z\to 0}\sum_k\sum_{k'} |V_{1|0k}|^2 |V_{1|0k'}|^2 
\frac1{z-\omega_k+\omega_{k'}}
\frac1{-\omega_1+\omega_0-\omega_k}\nonumber\\ &\times&
\left(\frac1{-\omega_1+\omega_0-\omega_k} 
+\frac1{\omega_1-\omega_0+\omega_{k'}}\right)\nonumber\\ 
&=&\lim_{z\to 0}\sum_k\sum_{k'} |V_{1|0k}|^2 |V_{1|0k'}|^2 
\frac{\omega_{k'}-\omega_{k}}{z-\omega_k+\omega_{k'}}
\frac1{-\omega_1+\omega_0-\omega_k}\nonumber\\ &\times&
\frac1{-\omega_1+\omega_0-\omega_k} 
\frac1{\omega_1-\omega_0+\omega_{k'}}\nonumber\\
&=&\sum_k\sum_{k'} |V_{1|0k}|^2 |V_{1|0k'}|^2 
\frac1{-\omega_1+\omega_0-\omega_k}
\frac1{-\omega_1+\omega_0-\omega_k} 
\frac1{\omega_1-\omega_0+\omega_{k'}}
\nonumber\\ && 
\label{A1.4}
\end{eqnarray}
The limit $z\to 0$ has become harmless: the resonant factor plays no role
since the fraction has a well defined limit.
Therefore, we have anew the cancellation of $C_{34}(0)$ with $C_{3'4'}(0)$.

We now turn to the contributions that involve 
the connecting vertex $\bar  L'_V$. 
They correspond to the following successions
that involve as last vertex (extreme left)
a change of the second index:
\newline
\noindent
(00;0k1;0kk'0;1k0;00), (00;0k1;0kk'0;1k'0;00),
(00;0k1;1k1k';1k0;00), 
\newline
\noindent
(00;0k1;1k1k';01k';00).
The corresponding contributions are called 
$C_5(z)$, $C_6(z)$, $C_7(z)$, $C_8(z)$.
They are:
\begin{eqnarray}
C_5(z)&=&\sum_k\sum_{k'} |V_{1|0k}|^2 |V_{1|0k'}|^2
\frac1{z+\omega_1-\omega_0-\omega_k}\frac1{z-\omega_k-\omega_{k'}}
\nonumber\\&\times&
\frac1{z-\omega_1+\omega_0-\omega_k} ,\nonumber\\
C_6(z)&=&\sum_k\sum_{k'} |V_{1|0k}|^2 |V_{1|0k'}|^2
\frac1{z+\omega_1-\omega_0-\omega_k}\frac1{z-\omega_k-\omega_{k'}}
\nonumber\\&\times&
\frac1{z-\omega_1+\omega_0-\omega_{k'}} ,\nonumber\\ 
C_7(z)&=&\sum_k\sum_{k'} |V_{1|0k}|^2 |V_{1|0k'}|^2
\frac1{z+\omega_1-\omega_0-\omega_k}\frac1{z-\omega_k+\omega_{k'}}
\nonumber\\&\times&
\frac1{z-\omega_1+\omega_0-\omega_k} ,\nonumber\\ 
C_8(z)&=&\sum_k\sum_{k'} |V_{1|0k}|^2 |V_{1|0k'}|^2
\frac1{z+\omega_1-\omega_0-\omega_k}\frac1{z-\omega_k+\omega_{k'}}
\nonumber\\&\times&
\frac1{z+\omega_1-\omega_0+\omega_{k'}}.    
\nonumber\\ && 
\label{A1.5}
\end{eqnarray}
The contributions involving as last vertex
a change of the second index can be obtained form those expression 
by replacing in the propagators $\omega \to -\omega$ 
for all values of the indices.
The corresponding contributions are called 
$C_{5'}$, $C_{6'}$, $C_{7'}$, $C_{8'}$.
With respect to $C_1(z)$, $C_2(z)$, $C_3(z)$, $C_4(z)$ , we note that
in $C_5(z)$, $C_6(z)$, $C_7(z)$, $C_8(z)$ the first propagator is resonant.
Let us consider the sum $C_{56}(z)$ of $C_5(z)$ and $C_6(z)$ at the limit
$z \to 0$.
Anew, the limit is taken whenever it is harmless.
\begin{eqnarray}
C_{56}(0)&=&\lim_{z\to 0}\sum_k\sum_{k'} |V_{1|0k}|^2 |V_{1|0k'}|^2 
\frac1{z+\omega_1-\omega_0-\omega_k}
\frac1{-\omega_k-\omega_{k'}} \nonumber\\ &\times& 
\left(\frac1{-\omega_1+\omega_0-\omega_k}
+\frac1{-\omega_1+\omega_0-\omega_{k'}} \right). 
\label{A1.6}
\end{eqnarray}
If we add the contribution $C_{5'6'}(0)$, we get:
\begin{eqnarray}
C_{56}(0)+C_{5'6'}(0)&=&\lim_{z\to 0}\sum_k\sum_{k'} |V_{1|0k}|^2 |V_{1|0k'}|^2 
\nonumber\\ &\times& 
\left(\frac1{z+\omega_1-\omega_0-\omega_k} 
+\frac1{z-\omega_1+\omega_0+\omega_k}\right)
\frac1{-\omega_k-\omega_{k'}} \nonumber\\ &\times& 
\left(\frac1{-\omega_1+\omega_0-\omega_k}
+\frac1{-\omega_1+\omega_0-\omega_{k'}} \right) \nonumber\\&=&
-2\pi i\sum_k\sum_{k'} |V_{1|0k}|^2 |V_{1|0k'}|^2 
\delta(\omega_1-\omega_0-\omega_k) 
\frac1{-\omega_k-\omega_{k'}} \nonumber\\ &\times& 
\left(\frac1{-\omega_1+\omega_0-\omega_k}
+\frac1{-\omega_1+\omega_0-\omega_{k'}} \right)\nonumber\\&=&
-2\pi i\sum_k\sum_{k'} |V_{1|0k}|^2 |V_{1|0k'}|^2 
\delta(\omega_1-\omega_0-\omega_k)
\nonumber\\ &\times&  
\frac1{-\omega_1+\omega_0-\omega_{k'}} 
\left(\frac1{2(-\omega_1+\omega_0)}
+\frac1{-\omega_1+\omega_0-\omega_{k'}} \right).
\nonumber\\ &&
\label{A1.7}
\end{eqnarray}
We have used the Dirac delta function to simplify some propagators 
to get the last relation.

We now turn to the sum $C_{78}(z)$ of $C_7(z)$ and $C_8(z)$ at the limit
$z \to 0$.
Anew, the limit is taken whenever it is harmless. 
\begin{eqnarray}
C_{78}(0)&=&\lim_{z\to 0}\sum_k\sum_{k'} |V_{1|0k}|^2 |V_{1|0k'}|^2 
\frac1{z+\omega_1-\omega_0-\omega_k}
\frac1{z-\omega_k+\omega_{k'}} \nonumber\\ &\times& 
\left(\frac1{-\omega_1+\omega_0-\omega_k}
+\frac1{\omega_1-\omega_0+\omega_{k'}} \right)\nonumber\\
&=&\lim_{z\to 0}\sum_k\sum_{k'} |V_{1|0k}|^2 |V_{1|0k'}|^2 
\frac1{z+\omega_1-\omega_0-\omega_k}
\frac{-\omega_k+\omega_{k'}}{z-\omega_k+\omega_{k'}} \nonumber\\ &\times& 
\frac1{-\omega_1+\omega_0-\omega_k}
\frac1{\omega_1-\omega_0+\omega_{k'}}\nonumber\\
&=&\lim_{z\to 0}\sum_k\sum_{k'} |V_{1|0k}|^2 |V_{1|0k'}|^2 
\frac1{z+\omega_1-\omega_0-\omega_k}\nonumber\\ &\times& 
\frac1{-\omega_1+\omega_0-\omega_k}
\frac1{\omega_1-\omega_0+\omega_{k'}}. 
\label{A1.8}
\end{eqnarray}
If we add the contribution $C_{7'8'}(0)$, we get
\begin{eqnarray}
C_{78}(0)+C_{7'8'}(0)&=&\lim_{z\to 0}\sum_k\sum_{k'} |V_{1|0k}|^2 |V_{1|0k'}|^2 \nonumber\\
&\times& 
\left(\frac1{z+\omega_1-\omega_0-\omega_k}+ 
\frac1{z-\omega_1+\omega_0+\omega_k}\right)\nonumber\\&\times&
\frac1{-\omega_1+\omega_0-\omega_k}
\frac1{\omega_1-\omega_0+\omega_{k'}} \nonumber\\ 
&=&-2\pi i\sum_k\sum_{k'} |V_{1|0k}|^2 |V_{1|0k'}|^2 
\delta(\omega_1-\omega_0-\omega_k)\nonumber\\ &\times& 
\frac1{-\omega_1+\omega_0-\omega_k}
\frac1{\omega_1-\omega_0+\omega_{k'}}\nonumber\\
&=&-2\pi i\sum_k\sum_{k'} |V_{1|0k}|^2 |V_{1|0k'}|^2 
\delta(\omega_1-\omega_0-\omega_k)\nonumber\\ &\times& 
\frac1{2(-\omega_1+\omega_0)}
\frac1{\omega_1-\omega_0+\omega_{k'}}\nonumber\\&&
\label{A1.9}
\end{eqnarray}
If we combine the non-vanishing contributions, we are left with
\begin{eqnarray}
W_{00.00}^{(4)}(0)&=&-2\pi i\sum_k\sum_{k'} |V_{1|0k}|^2 |V_{1|0k'}|^2 
\delta(\omega_1-\omega_0-\omega_k)
\nonumber\\ &\times&
\frac1{(\omega_1-\omega_0+\omega_{k'})^2}
\label{A1.10}
\end{eqnarray}
that is the looked-after expression.
It is therefore manifest that $W_{00.00}^{(4)}(0)$ does not vanish.

\section{Elements of $\tilde{\bar \Sigma}$ between off diagonal-off diagonal atomic states}
\setcounter{equation}{0}

\def\theequation{
C.\arabic{equation}}

The matrix elements of the resolvent of $R$ can also be expressed 
in terms of the irreductible  collision operator $\psi$:
Only elements $\psi_{a\bar a.a\bar a}$ 
play a role while elements  $\psi_{a\bar a.\bar a a}$ and $\psi_{a\bar a.\bar a a}$ vanish.
We have the alternative forms with respect to (\ref{4.1}):
\begin{eqnarray}
R_{a\bar a.a\bar a}(z)
&=&\frac1{z-\omega_a+\omega_{\bar a}-\psi_{a\bar a.a\bar a}(z)}.
\label{4.3}
\end{eqnarray}
The analytical properties of $\psi_{a\bar a.a\bar a}(z)$ are determined
by comparing the two expressions  of $R_{a\bar a.a\bar a}(z)$  
and are directly connected with the analytical properties of the resolvents of $H$ and $-H$.

For the reduced formalism, the role of $\psi$ is fulfilled 
by the operator $W$ (\ref{2.15}), the two operators differing by $T$ (\ref{2.16}) .
The elements of $\bar R(z)$ are expressed
in terms of  these of $R(z)$ and $T(z)$ 
\begin{eqnarray}
\bar R_{a\bar a.a\bar a}(z)&=&R_{a\bar a.a\bar a}+R_{a\bar a.a\bar a}
T_{a\bar a.a\bar a}\bar R_{a\bar a.a\bar a}
+R_{a\bar a.a\bar a}T_{a\bar a.\bar a a}\bar R_{\bar a a.a\bar a},\nonumber\\
\bar R_{a\bar a.\bar a a}(z)&=&R_{a\bar a.a\bar a}T_{a\bar a.a\bar a}\bar R_{a\bar a.\bar a a}
+R_{a\bar a.a\bar a}T_{a\bar a.\bar a a}\bar R_{a\bar a.a\bar a},
\label{4.5}
\end{eqnarray}
A little algebra enables to solve this system of equations, using the relation (\ref{2.17}) for
$W_{ab.cd}(z)$:
\begin{eqnarray}
\bar R_{a\bar a.a\bar a}(z)&=&
\frac{z-\omega_{\bar a}+\omega_a-W_{\bar a a.\bar a a}}
{(z-\omega_a+\omega_{\bar a}-W_{a\bar a.a\bar a})
(z-\omega_{\bar a}+\omega_a-W_{\bar a a.\bar a a})-T_{a\bar a.\bar
a a}T_{\bar a a.a\bar a}},
\nonumber\\
\bar R_{\bar a a.a\bar a}(z)&=&
\frac{T_{\bar a a.a\bar a}}
{(z-\omega_a+\omega_{\bar a}-W_{a\bar a.a\bar a})
(z-\omega_{\bar a}+\omega_a-W_{\bar a a.\bar a a})-T_{a\bar a.\bar a a}T_{\bar a a.a\bar a}}.
\nonumber\\&&
\label{4.7}
\end{eqnarray}
This form shows that all elements share poles 
due to the common denominator
(we are interested only to these poles 
that go either to $\omega_1-\omega_0$ or $\omega_0-\omega_1$ as $V^2\to 0$, 
and not to the singular points 
arising from the numerators of (\ref{4.7})).
The two off-diagonal elements ($\bar R_{\bar a a.a\bar a}$ for $a=1$ and $a=0$)  
are also linked with the diagonal ones by (cf. (\ref{3.17}))
\begin{equation}
(z-\omega_{\bar a}+\omega_a-\psi_{\bar a a.\bar a a})\bar R_{\bar a a.a\bar a}=
-(z-\omega_a+\omega_{\bar a}-\psi_{a\bar a .a\bar a})\bar R_{a\bar a .a\bar a}+1
\label{4.8}
\end{equation}
These relations can be interpreted by considering the last apparition, in $\bar R_{\bar a a.a\bar a}$,
of an element of $\bar L'_V$, leading to a state without field particle (an odd number of  $\bar L'_V$
vertex has to be present), and its replacement by an element of $L_V$, leading to an even number of
vertex $\bar L'_V$ . The new expression can be placed in relation with contributions to $\bar R_{a\bar
a .a\bar a}$.  
A supplementary contribution of $\bar R_{a\bar a .a\bar a}$ has to be treated separately:
Those relations can be checked directly from (\ref{4.7}).

If we take into account the vanishing of the off diagonal elements of $\psi$ to be able to replace in
(\ref{4.7}) the remaining $T$'s by $W$, the common denominator $D(z)$ in (\ref{4.7})  can be written as:
\begin{eqnarray}
D(z)&=&z^2
-z(W_{10.10}+W_{01.01}) -(\omega_1-\omega_0)^2
+(\omega_1-\omega_0)(W_{01.01}-W_{10.10})
\nonumber\\&+&
W_{10.10}W_{01.01}-W_{10.01}W_{01.10}
\label{4.9}
\end{eqnarray}  
Using (\ref{3.1}-\ref{3.2}) the denominator $D(z)$ takes a simpler form:
\begin{eqnarray}
D(z)&=&z^2
-z(W_{10.10}+W_{01.01}) -(\omega_1-\omega_0)^2
\nonumber\\&
+&(\omega_1-\omega_0)(W_{01.01}-W_{10.10})
\label{4.10}
\end{eqnarray}  
The analysis of the poles of the diagonal-diagonal elements in Ref. \cite{dH03b} 
can be repeated here for the elements relative to the diplole moments:
$D^{-1}(z)$ has the two poles for the same values of $z$ for which  the elements
$R_{a\bar a .\bar a a}(z)$ (\ref{4.1}), for $a=1$ and $a=0$, are singular, 
i.e. for the values $z=\omega_1+\zeta-\omega_0+\bar \delta \equiv \delta_{10}$
and  $z=\omega_0+ \delta-\omega_1+\bar\zeta \equiv \delta_{01}$.
We will note respectively by ${\cal A}_{10}$ and ${\cal A}_{01}$
the residue of $D^{-1}(z)$ at those poles, 
with the obvious property ${\cal A}_{10}={\cal A}_{01}^*$.
The difference with the unreduced formalism is that those residues 
are no longer given by simple products of ${\cal A}_1$, ${\cal A}_0$
defined from the Green's functions associated with the Hamiltonian.
The elements of $\bar R(z)$ can now be written  more explicitly as
\begin{eqnarray}
\bar R_{a\bar a.a\bar a}(z)&=&
\frac{z-\omega_{\bar a}+\omega_a-W_{\bar a a.\bar a a}(z)} {D(z)},\nonumber\\
\bar R_{\bar a a.a\bar a}(z)&=&
\frac{W_{\bar a a.a\bar a}(z)}{D(z)}=-\frac{W_{a\bar a.a\bar a}(z)}{D(z)},
\label{4.11}
\end{eqnarray}
from which the elements of $\tilde{\bar \Sigma}(t)$ can now be computed.
We have formally
\begin{eqnarray}
\tilde{\bar  \Sigma}_{10.10}(t)&=&
e^{-i\delta_{10} t}\,{\cal A}_{10}
\left(2\omega_1+\zeta -2\omega_0-\bar \delta
-W_{01.01}(\omega_1+\zeta-\omega_0+\bar \delta)\right)\nonumber\\&+&  
e^{-i\delta_{01} t}\,{\cal A}_{01}
\left(-\bar \zeta +\delta
- W_{01.01}(-\omega_1-\bar\zeta+\omega_0+ \delta)\right),
\nonumber\\
\tilde{\bar \Sigma}_{01.01}(t)&=&
e^{-i\delta_{01} t}\,{\cal A}_{01}
\left(-2\omega_1-\bar \zeta +2\omega_0+\delta
- W_{10.10}(-\omega_1-\bar\zeta+\omega_0+ \delta)\right)
\nonumber\\&+&
e^{-i\delta_{10} t}\,{\cal A}_{10}
\left(\zeta -\bar \delta
-W_{10.10}(\omega_1+\zeta-\omega_0+\bar \delta)\right),
\nonumber\\
\tilde{\bar \Sigma}_{01.10}(t)&=&
-e^{-i\delta_{10} t}\,{\cal A}_{10}
W_{10.10}(\omega_1+\zeta-\omega_0+\bar \delta)\nonumber\\&  
-&e^{-i\delta_{01} t}\,{\cal A}_{01}
W_{10.10}(-\omega_1-\bar\zeta+\omega_0+ \delta), 
\nonumber\\ 
\tilde{\bar \Sigma}_{10.01}(t)&=&
-e^{-i\delta_{01} t}\,{\cal A}_{01}
 W_{01.01}(-\omega_1-\bar\zeta+\omega_0+ \delta)
\nonumber\\&-&
e^{-i\delta_{10} t}\,{\cal A}_{10}
W_{01.01}(\omega_1+\zeta-\omega_0+\bar \delta).
\label{4.12}
\end{eqnarray}
Those relations are symbolized in the main text as:
\begin{eqnarray}
\tilde{\bar \Sigma}_{a\bar a.a\bar a}(t)&=&e^{-i\delta_{10} t}\alpha_{a\bar a.a\bar a} 
+ e^{-i\delta_{01} t} \beta_{a\bar a.a\bar a},
\nonumber\\
\tilde{\bar \Sigma}_{\bar a a.a\bar a}(t)&=&e^{-i\delta_{10} t}\alpha_{\bar a a.a\bar a}
+ e^{-i\delta_{01} t} \beta_{\bar a a.a\bar a}.
\label{4.13aa}
\end{eqnarray}
where the values of $\alpha$ and $\beta$ can be read on the preceding equations (\ref{4.12}).

$\tilde{\bar A}$ is evaluated from the vacuum-vacuum elements of the operator $\tilde{\bar
\Sigma}(t)$  for the value $t=0$.
The corresponding elements are:
\begin{equation}
\tilde{\bar A}_{a\bar a.a\bar a}=\alpha_{a\bar a.a\bar a}+ \beta_{a\bar a.a\bar a},
\qquad
\tilde{\bar A}_{\bar a a.a\bar a}=\alpha_{\bar a a.a\bar a}+ \beta_{\bar a a.a\bar a},
\label{4.14}
\end{equation}
The elements of the inverse $\tilde{\bar A}^{-1}$
of the $\tilde{\bar A}$ operator
can also be computed.
This requires the inversion  of a two by two matrix.
The determinant $|\tilde{\bar A}_D|$ of that matrix  is:
\begin{eqnarray}
|\tilde{\bar A}_D|&=& (\alpha_{10.10}+ \beta_{10.10})
(\alpha_{01.01}+ \beta_{01.01})
\nonumber\\&-&
(\alpha_{01.10}+ \beta_{01.10})(\alpha_{10.01}+ \beta_{10.01}).
\label{4.16}
\end{eqnarray}  
We can compute partial contributions from (\ref{4.12})) and obtain  
\begin{eqnarray}
&&\alpha_{10.10} \alpha_{01.01}
-\alpha_{10.01} \alpha_{01.10}
= 
{\cal A}^2_{10}\left[2(\omega_1-\omega_0)(\zeta -\bar \delta)
\right.\nonumber\\ &-&\left.
2(\omega_1-\omega_0)W_{10.10}(\omega_1+\zeta-\omega_0+\bar \delta)
+(\zeta -\bar \delta)^2\right.\nonumber\\&&\left.
-(\zeta -\bar \delta)\left(W_{01.01}(\omega_1+\zeta-\omega_0+\bar \delta)
+W_{10.10}(\omega_1+\zeta-\omega_0+\bar \delta)\right)\right].
\nonumber\\&\label{4.17}
\end{eqnarray}
We know that $\delta_{10}$ satisfies the equation $D(\delta_{10})=0$,
where $D(z)$ is given by (\ref{4.10}).
Therefore, we have
\begin{eqnarray}
D(\delta_{10})&=&\delta_{10}^2
-\delta_{10}(W_{10.10}(\delta_{10})+W_{01.01}(\delta_{10})) -(\omega_1-\omega_0)^2
\nonumber\\&
+&(\omega_1-\omega_0)(W_{01.01}(\delta_{10})-W_{10.10}(\delta_{10}))=0.
\label{4.18}
\end{eqnarray}  
Introducing the explicit value for 
$\delta_{10}$ ($\delta_{10}=\omega_1+\zeta-\omega_0+\bar \delta$), this relation (\ref{4.18})
becomes
\begin{eqnarray}
&&(\zeta -\bar \delta)^2+2(\omega_1-\omega_0)(\zeta -\bar \delta)
-2(\omega_1-\omega_0)W_{10.10}(\delta_{10})
\nonumber\\&&
+(\zeta -\bar \delta)(W_{10.10}(\delta_{10})+W_{01.01}(\delta_{10})) 
=0.
\label{4.19}
\end{eqnarray}  
A direct comparison with (\ref{4.17}) leads to the relation
\begin{equation}
\alpha_{10.10} \alpha_{01.01}
-\alpha_{10.01} \alpha_{01.10}= 0,
\label{4.20}
\end{equation}
and a similar relation holds for the $\beta$'s.
\begin{equation}
\beta_{10.10} \beta_{01.01}
-\beta_{10.01} \beta_{01.10}= 0. 
\label{4.21}
\end{equation}
The determinant $|\tilde{\bar A}_D|$ takes the simplest form
\begin{equation}
|\tilde{\bar A}_D|= \alpha_{10.10}\beta_{01.01} +\alpha_{01.01}\beta_{10.10}
- \alpha_{01.10}\beta_{10.01}- \alpha_{10.01}\beta_{01.10}.
\label{4.22}
\end{equation}
A little algebra provides the explicit result 
\begin{eqnarray}
|\tilde{\bar A}_D|
&=&{\cal A}_{10}{\cal A}_{01} 
\left[
\left(2\omega_1+\zeta -2\omega_0-\bar \delta\right)
\left(-2\omega_1-\bar \zeta +2\omega_0+\delta \right)
\right.\nonumber\\& &\left. 
+ \left(\zeta -\bar \delta \right) \left(-\bar \zeta +\delta \right)
- 2(\omega_1-\omega_0)
\left(W_{10.10}(-\omega_1-\bar\zeta+\omega_0+ \delta)
\right.\right.\nonumber\\& &\left.\left.
-W_{01.01}(\omega_1+\zeta-\omega_0+\bar \delta) \right)
\right.\nonumber\\&& \left. 
-(\zeta -\bar \delta-\bar \zeta - \delta)
\left(W_{10.10}(-\omega_1-\bar\zeta+\omega_0+ \delta)
\right.\right.\nonumber\\& &\left.\left.
+W_{01.01}(\omega_1+\zeta-\omega_0+\bar \delta) \right)
\right],
\nonumber\\&& 
\label{4.24}
\end{eqnarray}
In a perturbation expansion, $|\tilde{\bar A}_D|$ starts as $-4(\omega_1-\omega_0)^2$.
The elements of the inverse $\tilde{\bar A}^{-1}$ are
\begin{eqnarray}
(\tilde{\bar A}^{-1})_{a\bar a .a\bar a}&=&\frac1{|\tilde{\bar A}_D|}
(\alpha_{\bar a a.\bar a a}+ \beta_{\bar a a.\bar a a}), 
\nonumber\\
(\tilde{\bar A}^{-1})_{\bar a a.a\bar a}&=&-\frac1{|\tilde{\bar A}_D|} 
(\alpha_{\bar a a.a\bar a}+\beta_{\bar a a.a\bar a}).
\label{4.25}
\end{eqnarray}

\section{Determination of $\chi$}
\setcounter{equation}{0}

\def\theequation{
D.\arabic{equation}}

Denoting by $|\chi|$ the determinant of the $\chi$ matrix,
we have from the inversion of a standard $2\times 2$ matrix
\begin{eqnarray}
(\chi^{-1})_{11.11}&=&\chi_{00.00}\frac1{|\chi|},\nonumber\\
(\chi^{-1})_{00.00}&=&\chi_{11.11} \frac1{|\chi|},\nonumber\\
(\chi^{-1})_{00.11}&=&-\chi_{00.11} \frac1{|\chi|} ,\nonumber\\
(\chi^{-1})_{11.00}&=&-\chi_{11.00}\frac1{|\chi|}.
\label{5.10}
\end{eqnarray}

Using the explicit value for the elements of $\tilde{\bar\Theta}$, 
we get for the diagonal and off diagonal elements of the relation 
($a=b$ and $a \not= b$):
\begin{eqnarray}
&&\chi_{11.11} (\chi^{-1})_{11.11}- \chi_{11.00} (\chi^{-1})_{11.11} 
= \frac{W_{11.11}(0)}{W_{11.11}(0)+W_{00.00}(0)},\nonumber\\
&&\chi_{00.11} (\chi^{-1})_{11.00}- \chi_{00.00} (\chi^{-1})_{11.00} 
= \frac{W_{00.00}(0)}{W_{11.11}(0)+W_{00.00}(0)},\nonumber\\
&&\chi_{11.11} (\chi^{-1})_{11.00}- \chi_{11.00} (\chi^{-1})_{11.00} 
=- \frac{W_{00.00}(0)}{W_{11.11}(0)+W_{00.00}(0)},\nonumber\\
&&\chi_{00.11} (\chi^{-1})_{11.11}- \chi_{00.00} (\chi^{-1})_{11.11} 
=- \frac{W_{11.11}(0)}{W_{11.11}(0)+W_{00.00}(0)}. 
\nonumber\\&&   
\label{5.12}
\end{eqnarray}
Introducing the form (\ref{5.10}) for the inverse of $\chi$
into these relations (\ref{5.12}), we get
\begin{eqnarray}
&&\chi_{11.11}\chi_{00.00} - \chi_{11.00}\chi_{00.00}  
=|\chi| \frac{W_{11.11}(0)}{W_{11.11}(0)+W_{00.00}(0)},\nonumber\\
&&-\chi_{00.11}\chi_{11.00}+ \chi_{00.00}\chi_{11.00}
=|\chi| \frac{W_{00.00}(0)}{W_{11.11}(0)+W_{00.00}(0)},\nonumber\\
&&-\chi_{11.11}\chi_{11.00}+ \chi_{11.00}\chi_{11.00} 
=-|\chi| \frac{W_{00.00}(0)}{W_{11.11}(0)+W_{00.00}(0)},\nonumber\\
&&\chi_{00.11}\chi_{00.00}- \chi_{00.00}\chi_{00.00} 
=-|\chi| \frac{W_{11.11}(0)}{W_{11.11}(0)+W_{00.00}(0)}.  
\label{5.13}
\end{eqnarray}
Introducing the form (\ref{5.10}) for the inverse of $\chi$ in (\ref{5.9}), we get 
\begin{eqnarray}
\chi_{00.00}-\chi_{00.11}&=& |\chi|,\qquad
\chi_{11.11}-\chi_{11.00}=|\chi| ,\nonumber\\
\chi_{11.11}+\chi_{00.11}&=&1 ,\qquad
\chi_{00.00}+\chi_{11.00}=1. 
\label{5.14}
\end{eqnarray}
These relations enable to express all elements of $\chi$ in terms of $|\chi|$:
\begin{eqnarray}
\chi_{11.11}=\chi_{00.00}&=&\frac12(1+|\chi|) \nonumber\\
\chi_{11.00}=\chi_{00.11}&=&\frac12(1-|\chi|). 
\label{5.15}
\end{eqnarray}
We have indeed:
\begin{equation}
\chi_{11.11}\chi_{00.00} -\chi_{11.00}\chi_{00.11}
=\frac14(1+|\chi|)^2 -\frac14(1-|\chi|)^2
=|\chi|.
\label{5.16}
\end{equation}
Introducing those values (\ref{5.15}) into (\ref{5.13}), we get two relations
(the third and fourth relations in (\ref{5.13}) provide trivially the same relation
as the first two ones)
\begin{eqnarray}
(1+|\chi|)^2-(1+|\chi|)(1-|\chi|)
&=&4|\chi| \frac{W_{11.11}(0)}{W_{11.11}(0)+W_{00.00}(0)},\nonumber\\
-(1-|\chi|)^2+(1+|\chi|)(1-|\chi|)
&=&4|\chi| \frac{W_{00.00}(0)}{W_{11.11}(0)+W_{00.00}(0)}.  
\label{5.17}
\end{eqnarray}
Let us add and substract the two relations (\ref{5.17}).
We then get:
\begin{eqnarray}
(1+|\chi|)^2-(1-|\chi|)^2
&=&4|\chi| ,\nonumber\\
(1+|\chi|)^2+(1-|\chi|)^2-2(1+|\chi|)(1-|\chi|)
&=&4|\chi| \frac{W_{11.11}(0)-W_{00.00}(0)}{W_{11.11}(0)+W_{00.00}(0)}.  
\nonumber\\&&
\label{5.18}
\end{eqnarray}
The first relation is trivially satisfied and we are left with
the condition
\begin{equation}
4 |\chi|^2
=4|\chi| \frac{W_{11.11}(0)-W_{00.00}(0)}{W_{11.11}(0)+W_{00.00}(0)},  
\label{5.19}
\end{equation}
therefore 
\begin{equation}
|\chi|
= \frac{W_{11.11}(0)-W_{00.00}(0)}{W_{11.11}(0)+W_{00.00}(0)}.  
\label{5.20}
\end{equation}
From that expression, we get the explicit values for the first serie of elements of $\chi$ reported in
$\S5$.

We now turn to the other elements $\chi_{a\bar a.b\bar b}$.
In the  expression  $\chi (\exp -i\Phi t )\chi^{-1}$ is  diagonal and we get
\begin{equation}
\left(e^{-i\tilde{\bar \Theta} t}\right)_{a\bar a.c\bar c}
= \sum_b\chi_{a\bar a.b\bar b} e^{-i\delta_{b\bar b} t}(\chi^{-1})_{b\bar b.c\bar c}.
\label{5.29}
\end{equation}
Direct comparison between (\ref{5.29}) and (\ref{5.31}) provides us the conditions:
\begin{eqnarray}
\chi_{a\bar a.10}(\chi^{-1})_{10.c\bar c}&=&
\sum_b\alpha_{a\bar a.b\bar b}\left(\tilde{\bar A}^{-1}\right)_{b\bar b.c\bar c}, \nonumber\\
\chi_{a\bar a.01}(\chi^{-1})_{01.c\bar c}&=&
\sum_b\beta_{a\bar a.b\bar b}\left(\tilde{\bar A}^{-1}\right)_{b\bar b.c\bar c}.
\label{5.32}
\end{eqnarray}
We introduce the value of the inverse of the operators 
$\chi$ and $\tilde{\bar A}$. The form of the inverse depends on their diagonal or off-diagonal
character and the two possible cases have to be distinguished
\begin{eqnarray}
\chi_{a\bar a.10}\chi_{01.01}&=&\frac{|\chi|}{|\tilde{\bar A}_D|}
\left(\alpha_{a\bar a.10} (\alpha_{01.01}+ \beta_{01.01})
-\alpha_{a\bar a.01}(\alpha_{01.10}+ \beta_{01.10})\right),\nonumber\\
\chi_{a\bar a.10}\chi_{10.01}&=&-\frac{|\chi|}{|\tilde{\bar A}_D|}
\left(-\alpha_{a\bar a.10} (\alpha_{10.01}+ \beta_{10.01})
+\alpha_{a\bar a.01}(\alpha_{10.10}+ \beta_{10.10})\right),\nonumber\\
\chi_{a\bar a.01}\chi_{01.10}&=&-\frac{|\chi|}{|\tilde{\bar A}_D|}
\left(\beta_{a\bar a.10} (\alpha_{01.01}+ \beta_{01.01})
-\beta_{a\bar a.01}(\alpha_{01.10}+ \beta_{01.10})\right),\nonumber\\
\chi_{a\bar a.01}\chi_{10.10}&=&\frac{|\chi|}{|\tilde{\bar A}_D|}
\left(-\beta_{a\bar a.10} (\alpha_{10.01}+ \beta_{10.01})
+\beta_{a\bar a.01}(\alpha_{10.10}+ \beta_{10.10})\right).
\nonumber\\&\label{5.33}
\end{eqnarray}
The last two equations provide us the same condition as the first two ones, 
if we take into account  the relations (\ref{4.20}-\ref{4.21}).
The first two conditions (\ref{5.33})
provide, for $a=1$ and $a=0$, four relations that are simplified 
using the relations (\ref{4.20}-\ref{4.21})
\begin{eqnarray}
\chi_{10.10}\chi_{01.01}&=&\frac{|\chi|}{|\tilde{\bar A}_D|}
\left(\alpha_{10.10} \beta_{01.01}
-\alpha_{10.01}\beta_{01.10}\right),\nonumber\\
\chi_{01.10}\chi_{01.01}&=&\frac{|\chi|}{|\tilde{\bar A}_D|}
\left(\alpha_{01.10} \beta_{01.01}
-\alpha_{01.01}\beta_{01.10}\right),\nonumber\\ 
\chi_{10.10}\chi_{10.01}&=&-\frac{|\chi|}{|\tilde{\bar A}_D|}
\left(-\alpha_{10.10} \beta_{10.01}
+\alpha_{10.01}\beta_{10.10}\right),\nonumber\\
\chi_{01.10}\chi_{10.01}&=&-\frac{|\chi|}{|\tilde{\bar A}_D|}
\left(-\alpha_{01.10} \beta_{10.01}
+\alpha_{01.01}\beta_{10.10}\right).
\label{5.37}
\end{eqnarray}
Hermiticity of $\tilde{\bar\rho}_{at}(t)$ implies the obvious property 
$\tilde{\bar\rho}_{10}(t)=\tilde{\bar\rho}^*_{01}(t)$
that we impose also on the matrix $\tilde{\bar\rho}^P_{at}$ and
the following conditions have to hold:
\begin{eqnarray}
\chi_{10.10}&=&\chi^*_{01.01},\qquad\qquad
\chi_{01.10}=\chi^*_{10.01}.
\label{5.39}
\end{eqnarray}
Therefore, we are looking for a solution to (\ref{5.37}) under a form that incorporates these
conditions:
\begin{equation}
\chi_{10.10}=xe^{i\varphi},\quad
\chi_{01.01}=xe^{-i\varphi},\quad
\chi_{10.01}=ye^{i\psi},\quad
\chi_{01.10}=ye^{-i\psi},  
\label{5.40}
\end{equation}
and we have
$|\chi|=x^2-y^2$.
Introducing those values into (\ref{5.37}), we get:
\begin{eqnarray}
x^2&=&\frac{x^2-y^2}{|\tilde{\bar A}_D|}
\left(\alpha_{10.10} \beta_{01.01}
-\alpha_{10.01}\beta_{01.10}\right)\nonumber\\
xye^{i(\varphi-\psi)}&=&\frac{x^2-y^2}{|\tilde{\bar A}_D|}
\left(\alpha_{01.10} \beta_{01.01}
-\alpha_{01.01}\beta_{01.10}\right)\nonumber\\ 
xye^{-i(\varphi-\psi)}&=&-\frac{x^2-y^2}{|\tilde{\bar A}_D|}
\left(-\alpha_{10.10} \beta_{10.01}
+\alpha_{10.01}\beta_{10.10}\right)\nonumber\\
y^2&=&-\frac{x^2-y^2}{|\tilde{\bar A}_D|}
\left(-\alpha_{01.10} \beta_{10.01}
+\alpha_{01.01}\beta_{10.10}\right)
\label{5.42}
\end{eqnarray}
The first and fourth relations (\ref{5.42}) are obviously identical 
when the expression of $|\tilde{\bar A}_D|$ from (\ref{4.22}) is taken into account: their substraction
provides an identity . The second and the third ones are complex conjugate of each other.
and  provide the value of the phase difference $(\varphi-\psi)$.
The first relation leads to
\begin{equation}
y^2\left(\alpha_{10.10} \beta_{01.01}
-\alpha_{10.01}\beta_{01.10}\right)  
=-
x^2( \alpha_{01.01}\beta_{10.10}
- \alpha_{01.10}\beta_{10.01})
\label{5.45}
\end{equation}
and determines $y$ in function of $x$.
A further relation is needed to fix  the value of $x$.
A remaining indetermination is not new inside the context of subdynamics.\cite{PGHR73}\
It does not affect the evolution equations but concerns the relation between $\tilde{\bar\rho}^P$ and
$\tilde{\bar\rho}$.
We can connect, for instance, the value of ${|\chi|}$ with that of 
${|\tilde{\bar A}_D|}$ in a ``usual way":
${|\chi|^2}=|\tilde{\bar A}_D|$ or introduce a  criterion that ensures the positivity of the density
operator. We will not elaborate here further on this point.

From the point of view of a perturbation expansion, 
the elements $x$, $y$,$\alpha_{10.10}$ and $ \beta_{01.01}$ take the value 1 for $V\to 0$, 
while the other elements behave as $V^2$.

The relation (\ref{5.45}) can be made more explicit by replacing the $\alpha$'s 
and the $\beta$'s by their formal value.
\begin{eqnarray} 
&&x^2
\left[(\zeta -\bar \delta)(\bar \zeta - \delta)
-(\zeta -\bar \delta)W_{01.01}(\delta_{01})
- (\bar \zeta - \delta)W_{10.10}(\delta_{10})\right)]
\nonumber\\ 
&=&-y^2 
\left[(2\omega_1+\zeta -2\omega_0-\bar \delta)
(-2\omega_1+\bar \zeta +2\omega_0- \delta)
\right.\nonumber\\&-&\left.
(2\omega_1+\zeta -2\omega_0-\bar \delta)W_{10.10}(\delta_{01})
-(-2\omega_1+\bar \zeta +2\omega_0- \delta)
W_{01.01}(\delta_{10})\right].
\nonumber\\&&
\label{5.47}
\end{eqnarray}

\section{The presence of an incident field}
\setcounter{equation}{0}

\def\theequation{
E.\arabic{equation}}

\subsection{A passive incident field}
To evaluate (\ref{6.1}),
the matrix elements of the resolvent 
$\tilde{\bar R}(z)$ are required in terms of irreductible operators, starting from the perturbation
expansion
\begin{equation}
\tilde{\bar R}_{a\lambda b.c\lambda d}(z)=
\sum_{n=0}^{\infty} \left(\tilde{\bar R}^0(z) 
\left[\tilde{\bar L}_V \tilde {\bar R}^0(z)\right]^n\right)_{a\lambda b.c\lambda d}
\label{6.2}
\end{equation}
and a similar expression for $\tilde{\bar R}_{ab\lambda .cd\lambda}(z)$.

The irreductible operators  
$W_{a\lambda b.c\lambda d}$  and $W_{ab\lambda .cd\lambda}$   
for the complete liouvillian  $\tilde{\bar L}$ are defined through (\ref{2.15}):
\begin{equation}
W_{a\lambda b.c\lambda d}(z)=
\sum_{n=0}^{\infty} \left( \left[\tilde{\bar L}_V 
\tilde{\bar R}^0(z)\right]^n 
\tilde{\bar L}_V\right)_{a\lambda b.c\lambda d \left(irr\right)}
\label{6.3}
\end{equation}
\begin{equation}
W_{ab\lambda .cd\lambda }(z)=
\sum_{n=0}^{\infty} \left( \left[\tilde{\bar L}_V 
\tilde{\bar R}^0(z)\right]^n 
\tilde{\bar L}_V\right)_{ab\lambda .cd\lambda \left(irr\right)}
\label{6.4}
\end{equation}
The irreductibility condition holds with respect to the vacuum chosen, namely
the states involving the atom and the incident or emitted photons.

Since the field line $\lambda$ is not involved in any vertex 
and plays no role in determining the irreductibility condition, 
we have the obvious property:
\begin{equation}
W_{a\lambda b.c\lambda d}(z)=W_{ab.cd}(z-\omega_\lambda ),\qquad
W_{ab\lambda .cd\lambda }(z)=W_{ab.cd}(z+\omega_\lambda ).
\label{6.5}
\end{equation}
Those elements enable to write a compact form for the 
relevant elements of the resolvent $\tilde{\bar R} $:
\begin{equation}
\tilde{\bar R}_{a\lambda b.c\lambda d}(z)=
\tilde{\bar R}_{ab.cd}(z-\omega_\lambda),\qquad
\tilde{\bar R}_{ab\lambda .cd\lambda}(z)=
\tilde{\bar R}_{ab.cd}(z+\omega_\lambda)
\label{6.6}
\end{equation}
The poles that have been considered for the computation of the atomic part of $\tilde{\bar
\Sigma}(t)$ are the poles at $z=0$, $z=\theta$, $z=\delta_{10}$, $z=\delta_{01}$
that are present in $\tilde{\bar R}_{ab.cd}(z)$.
These poles are now shifted in (6.6) by $\pm \omega_\lambda$ and their residue is the same.
We use the notations previously introduced:
\begin{equation}
\tilde{\bar \Sigma}_{a\lambda a.b\lambda b}(t)
=e^{-i\omega_\lambda t}\alpha_{aa.bb}
+e^{-i(\bar \theta+\omega_\lambda)  t}\beta_{aa.bb}
\label{6.7}
\end{equation}
The values of the $\alpha$'s and the $\beta$'s can be obtained 
by identification with formulae (\ref{3.12}), (\ref{3.15}), (\ref{3.19}), (\ref{3.20})
or they can be read in the expression (\ref{3.25}) for $\tilde{\bar A}$
in $\S$3.

For the off-diagonal elements, we have
\begin{equation}
\tilde{\bar \Sigma}_{a\lambda \bar a.b\lambda \bar b}(t)
=e^{-i(\delta_{10}+\omega_\lambda) t}\alpha_{a\bar a.b\bar b}
+e^{-i(\delta_{01}+\omega_\lambda)  t} \beta_{a\bar a.b\bar b}
\label{6.8}
\end{equation}
In a similar way, we have
\begin{equation}
\tilde{\bar \Sigma}_{aa\lambda .bb\lambda }(t)
=e^{i\omega_\lambda t}\alpha_{aa.bb}
+e^{-i(\bar \theta-\omega_\lambda)  t} \beta_{aa.bb}
\label{6.9}
\end{equation}
\begin{equation}
\tilde{\bar \Sigma}_{a\bar a\lambda .b\bar b\lambda }(t)
=e^{-i(\delta_{10}-\omega_\lambda) t}\alpha_{a\bar a.b\bar b}
+e^{-i(\delta_{01}-\omega_\lambda)  t} \beta_{a\bar a.b\bar b}
\label{6.10}
\end{equation}
 A direct identification is possible:
\begin{eqnarray}
\tilde{\bar A}_{a\lambda b.c\lambda d}&=&
\tilde{\bar A}_{ab.cd}\qquad
\tilde{\bar A}_{ab\lambda .cd\lambda }=
\tilde{\bar A}_{ab.cd}
\label{6.11}
\\
\left(\tilde{\bar\Theta} \tilde{\bar A}\right)_{a\lambda a.b\lambda b}&=& 
\omega_\lambda \alpha_{aa.bb}
+(\bar \theta+\omega_\lambda)\beta_{aa.bb}\nonumber\\ 
\left(\tilde{\bar\Theta} \tilde{\bar A}\right)_{a\lambda \bar a.b\lambda \bar b}&=& 
(\delta_{10}+\omega_\lambda) \alpha_{a\bar a.b\bar b}
+(\delta_{01}+\omega_\lambda)\beta_{a\bar a.b\bar b}\nonumber\\
\left(\tilde{\bar\Theta} \tilde{\bar A}\right)_{aa\lambda .bb\lambda }&=& 
-\omega_\lambda \alpha_{aa.bb}
+(\bar \theta-\omega_\lambda)\beta_{aa.bb}\nonumber\\ 
\left(\tilde{\bar\Theta} \tilde{\bar A}\right)_{a\bar a\lambda .b\bar b\lambda }&=& 
(\delta_{10}-\omega_\lambda) \alpha_{a\bar a.b\bar b}
+(\delta_{01}-\omega_\lambda)\beta_{a\bar a.b\bar b}
\label{6.12}
\end{eqnarray}
The inverse $\tilde{\bar A}^{-1}$ of the operator $\tilde{\bar A}$
is known from the previous section thanks to the identification (\ref{6.11}).

We have the obvious property ($I_{ab.cd}$ is 1 when $a=c$ and $b=d$ 
and vanishes for the other possibilities)
\begin{eqnarray}
\left(\tilde{\bar\Theta} \tilde{\bar A}\right)_{a\lambda b.c\lambda d}&=&
\omega_\lambda \tilde{\bar A}_{ab.cd}
+\left(\tilde{\bar\Theta} \tilde{\bar A}\right)_{ab.cd} \nonumber\\
\left(\tilde{\bar\Theta} \tilde{\bar A}\right)_{ab\lambda .cd\lambda }&=&
-\omega_\lambda \tilde{\bar A}_{ab.cd}
+\left(\tilde{\bar\Theta} \tilde{\bar A}\right)_{ab.cd} 
\label{6.13}
\end{eqnarray}
from which the relations (\ref{6.15}) can be deduced.

\subsection{One absorbed incident field line}
The perturbation expansion of the resolvent 
$\tilde{\bar R}(z)$ can be written as
\begin{equation}
\tilde{\bar R}_{ab.c\lambda d}(z)=
\sum_{n=0}^{\infty} \left(\tilde{\bar R}^0(z) 
\left[\tilde{\bar L}_V \tilde{\bar R}^0(z)\right]^n\right)_{ab.c\lambda d}.
\label{7.2}
\end{equation}
A similar expression holds also for $\tilde{\bar R}_{ab.cd\lambda}(z)$ but only one
kind of elements will be displayed in details.
$\tilde{\bar R}_{ab.c\lambda d}(z)$ has to be expressed  in terms of the irreductible operators
$W_{ab.c\lambda d}$ defined in (\ref{2.15}) for the complete liouvillian  $\tilde{\bar L}$:
\begin{equation}
W_{ab.c\lambda d}(z)=
\sum_{n=0}^{\infty} \left( \left[\tilde{\bar L}_V 
\tilde{\bar R}^0(z)\right]^n 
\tilde{\bar L}_V\right)_{ab.c\lambda d \left(irr\right)}
\label{7.3}
\end{equation}
Those elements enable to write a compact form for the 
relevant elements of the resolvent $\tilde{\bar R} $:
\begin{equation}
\tilde{\bar R}_{ab.c\lambda d}(z)=
\sum_{ef}\sum_{gh} \tilde{\bar R}_{ab.ef}(z)
W_{ef.g\lambda h}(z)
\tilde{\bar R}_{g\lambda h. c\lambda d}(z).
\label{7.5}
\end{equation}
Since we have (the line $\lambda$ is passive through the resolvent and  the property
(\ref{6.6}) is used to get
\begin{equation}
\tilde{\bar R}_{ab.c\lambda d}(z)=
\sum_{ef}\sum_{gh} \tilde{\bar R}_{ab.ef}(z)
W_{ef.g\lambda h}(z)
\tilde{\bar R}_{gh. cd}(z-\omega_{\lambda})
\label{7.7}
\end{equation}
As in $\S$3, in absence of photons, 
$\tilde{\bar R}_{ab.ef}(z)=\bar R_{ab.ef}(z)$
The elements of the resolvent in (\ref{7.7}) are therefore the same one's 
that have been studied previously.

The poles to be considered for the computation of $\tilde{\bar \Sigma}(t)$
are the poles at $z=0$, $z=\theta$, $z=\delta_{10}$, $z=\delta_{01}$
that are present in $\tilde{\bar R}_{ab.ef}(z)$ 
and $\tilde{\bar R}_{gh.cd}(z)$.
These poles are well defined, as we have seen in the preceding sections.
Moreover, we have seen in Ref. \cite{dH03b} that it is possible to write 
the elements of $\tilde{\bar R}$ as a sum of expressions 
such that each of them contains only one relevant pole.
The computation of the residue does not therefore present 
conceptual problems.
The residues can also be evaluated directly from the expression (\ref{7.7}).
\begin{eqnarray} 
\tilde{\bar \Sigma}_{aa.d\lambda \bar d}(t)&=&
\sum_{b,c} \alpha_{aa.bb}
W_{bb.c\lambda \bar c}(0)
\tilde{\bar R}_{c\bar c.d\bar d}(-\omega_{\lambda})\nonumber\\
&+&e^{-i\bar \theta t}\sum_{b,c} \beta_{aa.bb}
W_{bb.c\lambda \bar c}(\bar \theta)
\tilde{\bar R}_{c\bar c.d\bar d}(\bar \theta-\omega_{\lambda})\nonumber\\ 
&+&    e^{-i(\delta_{10}+\omega_{\lambda})t}
\sum_{b,c}  \tilde{\bar R}_{aa.bb}(\delta_{10}+\omega_{\lambda})
W_{bb.c\lambda \bar c}(\delta_{10}+\omega_{\lambda})
\alpha_{c\bar c.d\bar d}\nonumber\\
&+&  e^{-i(\delta_{01}+\omega_{\lambda})t}   
\sum_{b,c}  \tilde{\bar R}_{aa.bb}(\delta_{01}+\omega_{\lambda})
W_{bb.c\lambda \bar c}(\delta_{01}+\omega_{\lambda})
\beta_{c\bar c.d\bar d}
\nonumber\\&&
\label{7.9}
\end{eqnarray}
Let us make the following comments for the correct computation of 
$\tilde{\bar R}_{c\bar c.d\bar d}(-\omega_{\lambda})$,
$\tilde{\bar R}_{c\bar c.d\bar d}(\bar \theta-\omega_{\lambda})$
$\tilde{\bar R}_{aa.bb}(\delta_{10}+\omega_{\lambda})$
and the similar expressions.
Displaying  the two relevant poles and their residue, 
we can indeed write
\begin{eqnarray}
\tilde{\bar R}_{c\bar c.d\bar d}(z-\omega_{\lambda})&=&
\alpha_{c\bar c.d\bar d}\frac1{z-\omega_{\lambda}-\delta_{10}}+
\beta_{c\bar c.d\bar d}\frac1{z-\omega_{\lambda}-\delta_{01}}
+r_{c\bar c.d\bar d}(z-\omega_{\lambda})
\nonumber\\&&
\label{7.13}
\end{eqnarray}
where the remaining function $r_{c\bar c.d\bar d}(z-\omega_{\lambda})$ is regular.
When we take the residue at the point $z=0$ and $z=\bar \theta$,
we merely replace $z$ in that expression by its corresponding value.
\begin{eqnarray}
\tilde{\bar R}_{c\bar c.d\bar d}(-\omega_{\lambda})&=&
\alpha_{c\bar c.d\bar d}\frac1{-\omega_{\lambda}-\delta_{10}}+
\beta_{c\bar c.d\bar d}\frac1{-\omega_{\lambda}-\delta_{01}}
+r_{c\bar c.d\bar d}(-\omega_{\lambda})\nonumber\\
\tilde{\bar R}_{c\bar c.d\bar d}(\bar \theta-\omega_{\lambda})&=&
\alpha_{c\bar c.d\bar d}\frac1{\bar \theta-\omega_{\lambda}-\delta_{10}}+
\beta_{c\bar c.d\bar d}\frac1{\bar \theta-\omega_{\lambda}-\delta_{01}}
+r_{c\bar c.d\bar d}(\bar \theta-\omega_{\lambda})
\nonumber\\&&
\label{7.14}
\end{eqnarray}
Therefore, we do not have to consider a possible deferred analytically continuation
with respect to the integration variable $\omega_{\lambda}$.\cite{dHG03}\
For consistency, 
when taking the residue at $z=\delta_{10}+\omega_{\lambda}$,
we first write $\tilde{\bar R}_{aa.bb}(z)$ as: 
\begin{equation}
\tilde{\bar R}_{aa.bb}(z)=\alpha_{aa.bb}\frac1{z}
+\beta_{aa.bb}\frac1{z-\bar\theta} +r_{aa.bb}(z)
\label{7.15}
\end{equation}
and we have:
\begin{eqnarray}
\tilde{\bar R}_{aa.bb}(\delta_{10}+\omega_{\lambda})&=&
\alpha_{aa.bb}\frac1{\delta_{10}+\omega_{\lambda}}
\nonumber\\
&+&\beta_{aa.bb}\frac1{\delta_{10}+\omega_{\lambda}-\bar\theta} 
+r_{aa.bb}(\delta_{10}+\omega_{\lambda})
\label{7.16}
\end{eqnarray}
We have therefore by direct identification from (\ref{7.9}):
\begin{eqnarray} 
\tilde{\bar A}_{aa.d\lambda \bar d}&=&
\sum_{b,c} \alpha_{aa.bb}
W_{bb.c\lambda \bar c}(0)
\tilde{\bar R}_{c\bar c.d\bar d}(-\omega_{\lambda})\nonumber\\
&+&\sum_{b,c} \beta_{aa.bb}
W_{bb.c\lambda \bar c}(\bar \theta)
\tilde{\bar R}_{c\bar c.d\bar d}(\bar \theta-\omega_{\lambda})\nonumber\\ 
&+&  
\sum_{b,c}  \tilde{\bar R}_{aa.bb}(\delta_{10}+\omega_{\lambda})
W_{bb.c\lambda \bar c}(\delta_{10}+\omega_{\lambda})
\alpha_{c\bar c.d\bar d}\nonumber\\
&+&    
\sum_{b,c}  \tilde{\bar R}_{aa.bb}(\delta_{01}+\omega_{\lambda})
W_{bb.c\lambda \bar c}(\delta_{01}+\omega_{\lambda})
\beta_{c\bar c.d\bar d}
\label{7.17}
\end{eqnarray}
\begin{eqnarray} 
\left(\tilde{\bar\Theta} \tilde{\bar A}\right)_{aa.d\lambda\bar d}&=&
\bar \theta 
\sum_{b,c} \beta_{aa.bb}
W_{bb.c\lambda \bar c}(\bar \theta)
\tilde{\bar R}_{c\bar c.d\bar d}(\bar \theta-\omega_{\lambda})\nonumber\\ 
&+&(\delta_{10}+\omega_{\lambda})    
\sum_{b,c}  \tilde{\bar R}_{aa.bb}(\delta_{10}+\omega_{\lambda})
W_{bb.c\lambda \bar c}(\delta_{10}+\omega_{\lambda})
\alpha_{c\bar c.d\bar d}\nonumber\\
&+&(\delta_{01}+\omega_{\lambda})  
\sum_{b,c}  \tilde{\bar R}_{aa.bb}(\delta_{01}+\omega_{\lambda})
W_{bb.c\lambda \bar c}(\delta_{01}+\omega_{\lambda})
\beta_{c\bar c.d\bar d}
\nonumber\\ &&
\label{7.21}
\end{eqnarray}
The inverse $\tilde{\bar A}^{-1}$ of the operator $\tilde{\bar A}$
can be computed easily.
From $\tilde{\bar A}\tilde{\bar A}^{-1}=I$, we get
\begin{equation}
0=\left(\tilde{\bar A}\tilde{\bar A}^{-1}\right)_{aa.d\lambda \bar d}=
\sum_b\tilde{\bar A}_{aa.bb}(\tilde{\bar A}^{-1})_{bb.d\lambda \bar d} 
+\sum_c \tilde{\bar A}_{aa.c\lambda \bar c}
(\tilde{\bar A}^{-1})_{c\lambda \bar c.d\lambda \bar d}
\label{7.25}
\end{equation}
Since we have obviously (the field line $\lambda$ is purely passive
and that case has been treated in the previous subsection):
\begin{equation}
(\tilde{\bar A}^{-1})_{c\lambda \bar c.d\lambda \bar d}=
(\tilde{\bar A}^{-1})_{c\bar c.d\bar d}
\label{7.26}
\end{equation}
and since $\tilde{\bar A}_{c\bar c.d\bar d}$,
$(\tilde{\bar A}^{-1})_{c\bar c.d\bar d}$ are known from  $\S$4 
while $\tilde{\bar A}_{aa.bb}$, $(\tilde{\bar A}^{-1})_{aa.bb}$ 
are  known from $\S$3, we have
\begin{equation}
(\tilde{\bar A}^{-1})_{aa.d\lambda \bar d} =
-\sum_{b,c} (\tilde{\bar A}^{-1})_{aa.bb}  
\tilde{\bar A}_{bb.c\lambda \bar c}
(\tilde{\bar A}^{-1})_{c\bar c.d\bar d}
\label{7.27}
\end{equation}
Those expressions enable the computation of the following elements of $\Theta$:
$\tilde{\bar\Theta}_{aa.b\lambda\bar b}$, $\tilde{\bar\Theta}_{b\bar b.a\lambda a}$,
$\tilde{\bar\Theta}_{aa.b\bar b\lambda}$, $\tilde{\bar\Theta}_{b\bar b.aa\lambda }$.
We have indeed:
\begin{eqnarray}
\tilde{\bar\Theta}_{aa.b\lambda\bar b}&=&
\left((\tilde{\bar\Theta} \tilde{\bar A})\tilde{\bar A}^{-1}\right)_{aa.b\lambda\bar b}\nonumber\\
&=&\sum_c (\tilde{\bar\Theta} \tilde{\bar A})_{aa.cc} 
(\tilde{\bar A}^{-1})_{cc.b\lambda\bar b}+
\sum_c (\tilde{\bar\Theta} \tilde{\bar A})_{aa.c\lambda\bar c} 
(\tilde{\bar A}^{-1})_{c\lambda\bar c.b\lambda\bar b}
\nonumber\\&&
\nonumber\\&\label{7.31}
\end{eqnarray}
Using the expressions (\ref{7.25}) and (\ref{7.27}) for 
$(\tilde{\bar A}^{-1})_{c\lambda\bar c.b\lambda\bar b}$ 
and $(\tilde{\bar A}^{-1})_{cc.b\lambda\bar b}$, we get
\begin{eqnarray}
\tilde{\bar\Theta}_{aa.b\lambda\bar b}&=&
\sum_{c,d,e} (\tilde{\bar\Theta} \tilde{\bar A})_{aa.cc} 
(\tilde{\bar A}^{-1})_{cc.dd}  \tilde{\bar A}_{dd.e\lambda \bar e}
(\tilde{\bar A}^{-1})_{e\bar e.b\bar b}+
\sum_c (\tilde{\bar\Theta} \tilde{\bar A})_{aa.c\lambda\bar c} 
(\tilde{\bar A}^{-1})_{c\bar c.b\bar b}\nonumber\\&=&
\sum_{d,e} \tilde{\bar\Theta}_{aa.dd} 
\tilde{\bar A}_{dd.e\lambda \bar e}
(\tilde{\bar A}^{-1})_{e\bar e.b\bar b}+
\sum_c (\tilde{\bar\Theta} \tilde{\bar A})_{aa.c\lambda\bar c} 
(\tilde{\bar A}^{-1})_{c\bar c.b\bar b}
\label{7.32a}
\end{eqnarray}
The last result is reproduced in (\ref{7.32}).


\begin{thebibliography}{99}

\bibitem{dH03a}
M.~de~Haan,
\newblock {Annals of Physics, {\bf311} (2004), 314.} 


\bibitem{dHG00a}
M.~de~Haan and C.~George, Trends in Statistical Physics {\bf 3} (2000), 115.

\bibitem{dHG03}
M.~de~Haan, and C.~George,
\newblock {\em Progr. Theor. Phys.} {\bf 109} (2003), 881.

\bibitem{dH91}
M.~de~Haan, Physica {\bf A171} (1991), 159.


\bibitem{CT94}
C.~Cohen-Tannoudji.
\newblock {\em Atoms in Electromagnetic Fields}.
\newblock {World scientific} (1994), paper 2.1, p119. 

\bibitem{Mo75}
B.~R.~Mollow, Phys. Rev. {\bf 12} (1975), 1919.

\bibitem{PT97}
E.~A.~Power and T.~Thirunamachandran,
 Phys. Rev.A {\bf 56}  (1997), 3395.

\bibitem{dH85}
M.~de~Haan, Physica {\bf A132} (1985), 375 and 397.



\bibitem{Bal75}
R.~Balescu, {\it Equilibrium and Nonequilibrium Statistical Mechanics}
(Wiley-Interscience, New York, 1975).


\bibitem{NN58}
N.~Nakanishi,
\newblock {\em Progr. Theor. Phys.} {\bf 19} (1958), 607.

\bibitem{dHH73a}
M.~de~Haan and F.~Henin,
Physica {\bf 67} (1973), 197.

\bibitem{dHG04}
M.~de~Haan and C.~George,
Bull. Cl. Sc. Ac. Roy. Bel. {\bf 1-6} (2004).


\bibitem{PGH69}
I.~Prigogine, C.~George and F.~Henin,
 Physica {\bf 45} (1969), 418.

\bibitem{PGHR73}
I.~Prigogine, C.~George, F.~Henin and L.~Rosenfeld, Chemica Scripta {\bf 4} (1973), 5.

\bibitem{dH98}
M.~de~Haan,
Bull. Cl. Sc. Ac. Roy. Bel. {\bf 1-6} (1998), 111.


\bibitem{OPP01}
G.~Ordonez, T.~Petrosky and I.~Prigogine,
 Phys. Rev.A {\bf 63}  (2001), 052106.

\bibitem{dH04b}
M.~de~Haan,
\newblock{lanl.arXiv:physics/0405023},
\newblock {presented Annals of Physics  (2004).} 


\bibitem{CP92}
P.~Coveney and O.~Penrose.
\newblock {\em J. Phys. A Math. Gen.} {\bf25} (1992), L4947.


\bibitem{dH03b}
M.~de~Haan, 
Bull. Cl. Sc. Ac. Roy. Bel. {\bf 7-12}, (2003), 267.

\bibitem{SCG78}
E.~C.~G.~Sudarshan, C.~B.~Chiu and V.~Gorini,
Phys. Rev.D {\bf 18}  (1978), 2914.

\bibitem{BP74}
R.~Balescu and M.~Poulain,
 Physica {\bf 76} (1974), 421.


\bibitem{dHG00b}
M.~de~Haan and C.~George,
Bull. Cl. Sc. Ac. Roy. Bel. 6, {\bf1-6} (2000), 129.


\bibitem{dHG02}
M.~de~Haan and C.~George,
Bull. Cl. Sc. Ac. Roy. Bel. {\bf1-6} (2002), 9.


\bibitem{ELOP66}
R.~Eden, P.~Landshoff, D.~Olive and J.C.~Polkinghorne.
\newblock {\em The Analytic S Matrix}.
\newblock {Cambridge University Press}, 1966.


\bibitem{GH74}
G. C. Hegerfelt, 
 Phys. Rev.D {\bf 10}  (1974), 3320.

\bibitem{HR79}
G. C. Hegerfelt and S. N. Ruiijsenaars, 
 Phys. Rev.D {\bf 22}  (1979), 377.

\bibitem{GH94}
G. C. Hegerfelt, 
 Phys. Rev. Lett. {\bf 72}  (1994), 596.


\bibitem{GH98}
G. C. Hegerfelt,
\newblock {\em Irreversibility and Causality in Quantum Theory-Semigroups and Rigged Hilbert
Spaces}, ed. by A. Bohm, H.-D. Dobner and P. Kielanovski,
\newblock {Springer Lectures Notes {\bf 504}}, 1998.

\bibitem{MJF95}
P.~W.~Milonni, D.~F.~V~James and H.~Fearn,
 Phys. Rev.A {\bf 52}  (1995), 1525.

\bibitem{SK30}
 S. ~Kikuchi,  
 Zeit. Phys. {\bf 66}  (1930), 558. 

\bibitem{EF32}
 E.~Fermi  
Rev. Mod. Phys. {\bf 4} (1932) ,87.  

\bibitem{GB33}
 G.~Breit, 
Rev. Mod. Phys. {\bf 5} (1933), 91.  

\bibitem{MK74}
 P. W.~Milonni and P. L. Knight,
 Phys. Rev.A {\bf 10}  (1974), 1096. 

\bibitem{BCPPP90}
A.~K.~Biswas, G.~Compagno, G.~M.~Palma, R.~Passante and F.~Persico,
 Phys. Rev.A {\bf 42}  (1990), 4291.

\bibitem{CPPP95}
G.~Compagno, G.~M.~Palma, R.~Passante and F.~Persico,
Chem. Phys. {\bf 198}  (1995), 19.

\bibitem{RE09} 
W. Ritz and A. Einstein,
Phys. Zeit. {\bf10} (1909), 323.

\bibitem{POP01}
T.~Petrosky, G.~Ordonez and I.~Prigogine,
 Phys. Rev.A {\bf 64}  (2001), 062101.

\bibitem{RB04}
R.~C.~Bishop,
Studies in History and Philosophy of Modern Physics {\bf 35}  (2004), 1.

\end{thebibliography}
\end{document}